\documentclass[aps,prl,nobibnotes,twocolumn,superscriptaddress,bibliography]{revtex4-2}

\usepackage{amsfonts}
\usepackage{mathrsfs}
\usepackage{amsmath}
\usepackage{color}
\usepackage{graphicx}
\usepackage{bm}
\usepackage{amssymb}
\usepackage{xspace}
\usepackage{epstopdf}
\usepackage{dcolumn}
\usepackage{longtable}
\usepackage{multirow}
\usepackage{float}
\usepackage{comment}

\usepackage[colorlinks=true, letterpaper=true, pdfstartview=FitV, linkcolor=blue, citecolor=blue, urlcolor=blue]{hyperref}

\makeatother

\begin{document}
\title{Observation of Single Pair of Type-III Weyl Points in Sonic Crystals }
\author{Xiao-Ping Li}
\thanks{These authors contributed equally to this work.}
\affiliation{Centre for Quantum Physics, Key Laboratory of Advanced Optoelectronic Quantum Architecture and Measurement (MOE), School of Physics, Beijing Institute of Technology, Beijing, 100081, China}
\affiliation{Beijing Key Lab of Nanophotonics \& Ultrafine Optoelectronic Systems, School of Physics, Beijing Institute of Technology, Beijing, 100081, China}
\author{Feng Li}
\thanks{These authors contributed equally to this work.}
\affiliation{Centre for Quantum Physics, Key Laboratory of Advanced Optoelectronic Quantum Architecture and Measurement (MOE), School of Physics, Beijing Institute of Technology, Beijing, 100081, China}
\affiliation{Beijing Key Lab of Nanophotonics \& Ultrafine Optoelectronic Systems, School of Physics, Beijing Institute of Technology, Beijing, 100081, China}
\author{Di Zhou}
\affiliation{Centre for Quantum Physics, Key Laboratory of Advanced Optoelectronic Quantum Architecture and Measurement (MOE), School of Physics, Beijing Institute of Technology, Beijing, 100081, China}
\affiliation{Beijing Key Lab of Nanophotonics \& Ultrafine Optoelectronic Systems, School of Physics, Beijing Institute of Technology, Beijing, 100081, China}
\author{Ying Wu}
\email{wuyinghit@gmail.com}
\affiliation{School of Physics and Optoelectronics, South China University of Technology, Guangzhou, Guangdong 510640, China}
\author{Zhi-Ming Yu}
\email{zhiming\_yu@bit.edu.cn}
\affiliation{Centre for Quantum Physics, Key Laboratory of Advanced Optoelectronic Quantum Architecture and Measurement (MOE), School of Physics, Beijing Institute of Technology, Beijing, 100081, China}
\affiliation{Beijing Key Lab of Nanophotonics \& Ultrafine Optoelectronic Systems, School of Physics, Beijing Institute of Technology, Beijing, 100081, China}
\author{Yugui Yao}
\email{ygyao@bit.edu.cn}
\affiliation{Centre for Quantum Physics, Key Laboratory of Advanced Optoelectronic Quantum Architecture and Measurement (MOE), School of Physics, Beijing Institute of Technology, Beijing, 100081, China}
\affiliation{Beijing Key Lab of Nanophotonics \& Ultrafine Optoelectronic Systems, School of Physics, Beijing Institute of Technology, Beijing, 100081, China}

\begin{abstract}
In electronics systems, the Weyl points can be classified into three types based on the geometry of the Fermi surface, and each type exhibits various unique and intriguing phenomena. While the type-I and type-II Weyl points have been achieved in both spinful and spinless systems, the realization of type-III Weyl points remains challenging, and has not been reported in artificial periodic systems. Here, we for the first time report the experimental observation of the type-III Weyl points in a sonic crystal. Remarkably, a single pair of type-III Weyl points are observed as the only band crossings in a frequency range, experimentally disproving a common belief in the field, namely, the minimal number of Weyl points in nonmagnetic systems is four. The consistency between experimental results and theoretical predictions confirms the existence of type-III Weyl points, noncontractible Fermi arc surface states, and chiral edge states. Our work not only fill the gap of the type-III Weyl point in sonic crystal but also stimulate related researches in other systems, such as photonic, mechanical and cold atom systems.
\end{abstract}
\maketitle

\textit{\textcolor{blue}{Introduction}}\textit{.} Quasiparticles, such as Weyl/Dirac fermion, have attracted much attention because they can be manifested by the low energy excitation in Weyl/Dirac semimetals \cite{weylDirac,murakami2007phase,wan2011topological,weng2015weyl,yu2022encyclopedia}. In a Weyl semimetal, the conduction bands and valence bands can cross linearly and produce twofold-degenerate Weyl points with divergent Berry curvature and quantized topological charges, which result in many extraordinary properties, such as chiral magnetic effect \cite{burkov2015chiral}, quantized circulation of anomalous shift \cite{liu2020quantized}, and other unusual optical and magnetic responses  \cite{nagaosa2020transport,de2017quantized,zhou2015plasmon}. Linear-crossing Weyl points carrying a topological charge of $\mathcal{C}=\pm1$ can exist without any symmetry except translational symmetry. In solids, however, there exist rich crystal symmetries that gives birth to the Weyl point with non-linear dispersion, as well as larger topological charges, i.e., charge-2 Weyl point ($\mathcal{C}=\pm2$), charge-3 Weyl point ($\mathcal{C}=\pm3$) and charge-4 Weyl point ($\mathcal{C}=\pm4$)  \cite{yu2022encyclopedia,fang2012multi,zhang2020twofold,cui2021charge}. These unconventional Weyl points can exhibit more novel phenomena due to higher topological charge \cite{nagaosa2020transport,cui2021charge}. Moreover, the non-linear energy dispersions can also lead to striking non-Fermi-liquid behaviors \cite{han2019emergent,wang2019topological,zhang2021quantum}.

\begin{figure}[t]
\includegraphics[width=8.2cm]{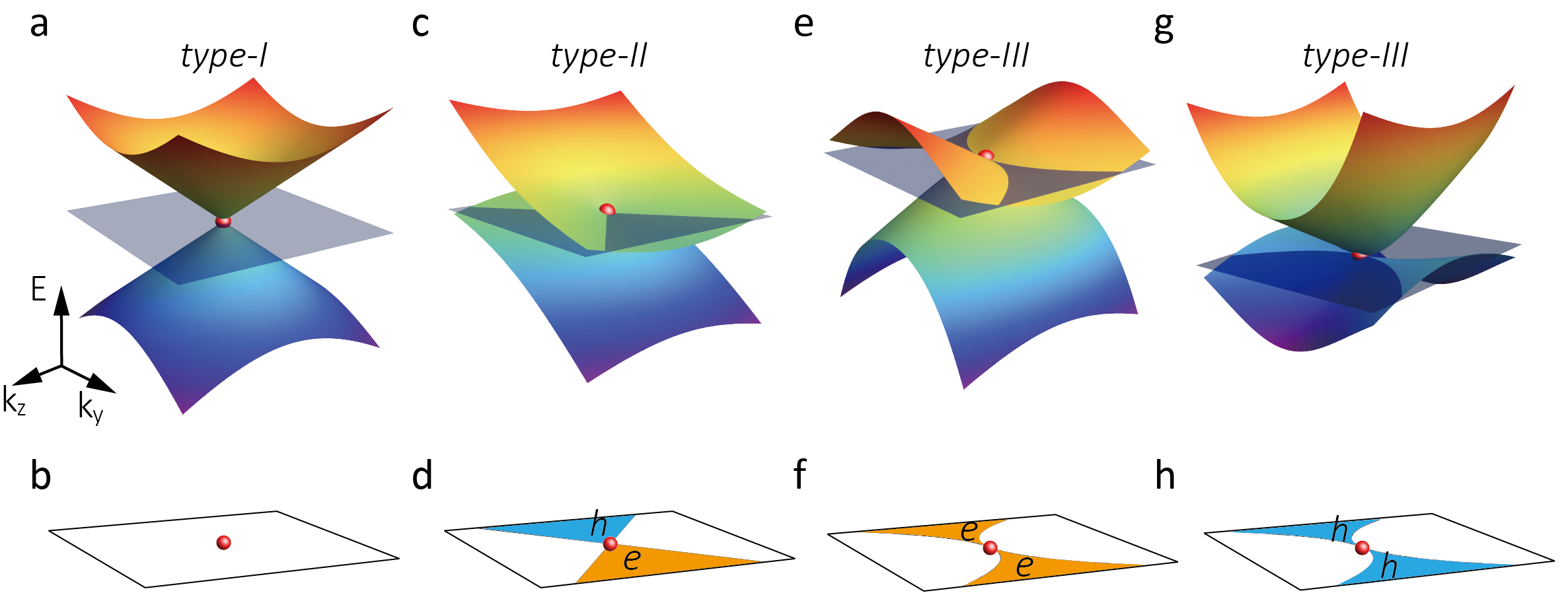} \caption{Schematic illustration of three types of the Weyl points.  (a,b) Type-I Weyl cone with point-like Fermi surface. (c,d) The Fermi surface of a type-II Weyl cone includes both electron and hole pockets, which are denoted as yellow and blue in (d). (e,f) A type-III Weyl point is the contact point between electron and electron pockets or between hole and hole pockets. Notice that type-III Weyl points must have higher-order (quadratic or cubic)  band dispersion and higher Chern number.  The grey plane denotes the Fermi level.
\label{fig1}}
\end{figure}

Besides the classification based on topological charge, the Weyl points also can be classified with respect to the geometry of their Fermi surfaces, as proposed by A. A. Soluyanov \emph{et. al} in Ref. \cite{soluyanov2015type}, where the conventional linear Weyl point is classified into two different types. The type-I Weyl point hosts a point-like Fermi surface [see Fig.~\ref{fig1}(a), (b)], and the type-II hosts an electron-hole pocket Fermi surface [Fig.~\ref{fig1}(c), (d)] due to linearly tilted bands. However, Li et al \cite{li2021type} recently found that for the unconventional Weyl points with non-linear dispersion, there exists another possibility of Fermi surface, which is constructed by two touched electrons (or hole) pockets [see Fig.~\ref{fig1}(e), (f) or Fig.~\ref{fig1}(g), (h)], due to the quadratically tilted bands, and term such types of Weyl point as type III.
The type-III Fermi surface cannot be realized in the conventional charge-1 Weyl point, and can only be found in the Weyl points with higher topological charge. Fig.~\ref{fig1} illustrates the characteristic band dispersion of the three types of Weyl points. For type-I Weyl point, the slopes of the two bands have opposite sign along any $k$ direction [see Fig.~\ref{fig1}(a)], while for type-II Weyl point, the slopes can have same sign along certain $k$ directions, as shown in Fig.~\ref{fig1}(c). The type-III Weyl point can be a realization in charge-2 Weyl system, which has band dispersion of parabola with the same opening direction along one certain $k$ direction, as shown in Fig.~\ref{fig1}(e) and Fig.~\ref{fig1}(g). Since most of the low-energy behaviors of systems in solids are determined by the geometry of the Fermi surface, the type-II and type-III Weyl point can exhibit many novel distinctive physical properties, such as magnetic breakdown \cite{o2016magnetic}, Landau level collapse  \cite{yu2016predicted,tchoumakov2016magnetic,udagawa2016field} and other unusual magnetoresponses \cite{yu2016predicted}.
The type-I and type-II conventional charge-1 Weyl points, and type-I charge-2 Weyl points have been reported in  electronic systems \cite{lv2021experimental,lv2015experimental,deng2016experimental,huang2016spectroscopic,yao2019observation,xu2017discovery,chang2016strongly,belopolski2017signatures,huang2016new,takane2019observation} and artificial periodic structures, such as photonic crystals \cite{lu2014topological,shastri2017realizing,yang2018ideal,saba2017group,yang2020ideal}, sonic crystals \cite{li2018weyl,ge2018experimental,xie2019experimental,huang2020ideal,he2020observation,yang2019topological,he2018topological,peri2019axial}, and electric circuits \cite{lee2018topolectrical,li2021ideal}.

So far, the type-III Weyl point has been verified in only one electronic material, which however would be gapped by charge density wave effect at low temperature \cite{li2021type}. Thus, it is crucial to find ideal platforms hosting type-III Weyl points. Here, the condition for being ``ideal" is that the system should have only a single pair of type-III Weyl points in the Brillouin zone (BZ) and the two bands forming the Weyl points are isolated and well separated from other bands, such that the signal will be entirely caused by the type-III Weyl points, without interference from other band crossings and trivial bands.  This is a challenging task and has not been achieved for type-III Weyl points and other (doubly degenerate) Weyl points. Particularly, there exists a common belief in the field of topological semimetals that the minimal number of Weyl points in nonmagnetic systems is four. Fortunately, it was recently shown while the above argument is valid for spinful (electronic) systems, but not for spinless systems \cite{wang2022single}.

In this work, we filled in the gaps by reporting on the experimental observation of ideal type-III Weyl points state in a three-dimensional sonic crystal, name, only a single pair of type-III Weyl points formed by the lowest two bands, and these two bands are well separated from other bands. First, we construct a two-band tight-binding (TB) model in a tetragonal lattice to show how the idea originates. Based on the model constructing and symmetry analysis, we fabricate a sonic crystal, which contains a pair of type-III Weyl points that emerge at time-reversal invariant momenta protected by the $D_{4}$ point group. Especially, the frequencies of the two bands associated with the type-III Weyl point are in the range of 0-5.3 kHz, and no other trivial bands occupying this frequency range, which is beneficial for the potential applications of topological acoustic devices. Both simulated and experimental results confirm the characteristic dispersion of the type-III Weyl points. In addition, as a signature of the Weyl phase, the surface Fermi arc has also been observed in our system.
The surface Fermi arc for the single pair of Weyl points  state is extended and exhibits a non-contractible winding topology on the surface BZ torus.

\textit{\textcolor{blue}{Rationale and lattice model}}\textit{.}
To realize type-III Weyl points in sonic systems, we need to construct a spinless crystal with unconventional charge-$n$ ($n>1$) Weyl points and the band dispersion of the unconventional Weyl points is quadratically tilted.
The former condition can be inferred from crystalline symmetry, while the latter requires fine-tuning, which can be easily achieved in sonic crystals.
As shown in Ref. \cite{yu2022encyclopedia}, a charge-2 Weyl point can appear in space group (SG) No. 89 as an essential spinless excitation located at one of four high-symmetry points ($\Gamma$, $Z$, $M$, $A$).
Thus, we construct a  lattice model to investigate the possibility of type-III Weyl points in SG 89.

\begin{figure*}[t]
\includegraphics[width=15.2cm]{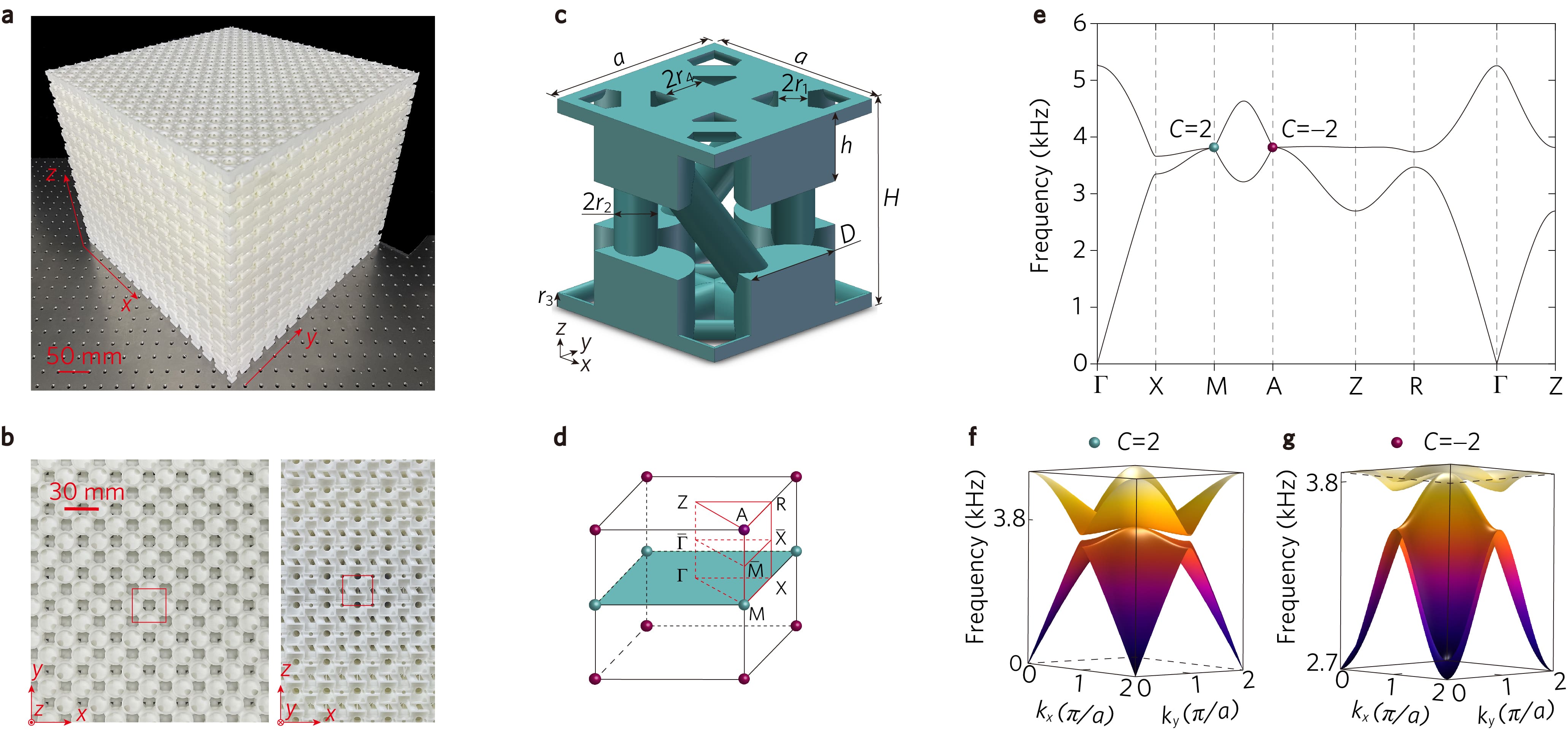} \caption{The sonic crystal structure of the single pair of type-III Weyl point states and its bulk band structures. (a) A photo of the experimental sample. (b) Enlarged view of sample on top and side view. A unit cell is outlined with red squares. (c) Schematic diagram of the geometry structure of the unit cell. (d) The bulk BZ along with the distribution of the Weyl points. The Weyl points with charges $-2$ and $+2$ are represented by purple and cyan spheres, respectively. (e) The simulated band structure of the acoustic lattice.  (f)-(g) Band structures  of the type-III Weyl point at $M$ and $A$ in the $k_{x}$-$k_{y}$ plane.
\label{fig2}}
\end{figure*}

A symmetry-allowed two-band TB model of SG 89 is given by \cite{sm}
\begin{eqnarray}
\mathcal{H}	=	h_{x}\sigma_{x}+h_{y}\sigma_{y}+h_{z}\sigma_{z}+h_{0}\sigma_{0},\label{eq:hamt3}
\end{eqnarray}
with $\sigma_{0}$ an $2\times2$ identity matrix and $\sigma_{i}$ ($i$=$x$, $y$, $z$)  the Pauli matrices. Here,
$h_{x}=4\,\mathrm{cos}\,\frac{k_{x}}{2}\,\mathrm{cos}\,\frac{k_{y}}{2}\,(t_{1}+t_{2}\,\mathrm{cos}\,k_{z})$, $h_{y}=4\,t_{2}\,\mathrm{sin}\,\frac{k_{x}}{2}\,\mathrm{sin}\,\frac{k_{y}}{2}\,\mathrm{sin}\,k_{z}$, $h_{z}=2\,(t_{3}-t_{4})\,(\mathrm{cos}\,k_{x}-\mathrm{cos}\,k_{y})$ and $h_{0}=2\,(t_{3}+t_{4})\,(\mathrm{cos}\,k_{x}+\mathrm{cos}\,k_{y})$ with $t$'s real parameters.
One can easily find that the two bands will degenerate at $M$ ($\pi, \pi, 0$) and $A$ ($\pi, \pi, \pi$) points, forming two Weyl points.
To further study the low energy features of the Weyl points, we expand the TB Hamiltonian around $M$ and $A$ points. The corresponding effective Hamiltonians, to the leading order, read
\begin{eqnarray}
H_{M} & = & (t_{1}+t_{2})k_{x}k_{y}\sigma_{x}+4t_{2}k_{z}\sigma_{y}+(t_{3}-t_{4})(k_{x}^{2}-k_{y}^{2})\sigma_{z}\nonumber\\
 &  & +(t_{3}+t_{4})(k_{x}^{2}+k_{y}^{2})\sigma_{0}
,\label{kpM}
\end{eqnarray}
for $M$ point and
\begin{eqnarray}
H_{A} & = & (t_{1}-t_{2})k_{x}k_{y}\sigma_{x}-4t_{2}k_{z}\sigma_{y}+(t_{3}-t_{4})(k_{x}^{2}-k_{y}^{2})\sigma_{z}\nonumber\\
 &  & +(t_{3}+t_{4})(k_{x}^{2}+k_{y}^{2})\sigma_{0}
,\label{kpA}
\end{eqnarray}
for $A$ point.
From  Eqs. (\ref{kpM}) and (\ref{kpA}), one knows the Weyl points at $M$ and $A$ points are charge-2 Weyl points with linear dispersion along $k_z$ direction  and quadratic dispersion in $k_x$-$k_y$ plane, enforced by the two-dimensional irreducible representation of the $D_{4}$ point group \cite{yu2022encyclopedia}.

Interestingly, the last term in Eqs. (\ref{kpM}-\ref{kpA}) with $\sigma_{0}$ matrix denotes the tilt of spectrum in $k_x$-$k_y$ plane, which is a key to realize the type-III Weyl points \cite{li2021type}.
The Weyl point at both M and A points become type III when  $\left|t_{3}+t_{4}\right|>\left|t_{3}-t_{4}\right|$, as in such case both bands (parabolas) along $k_x$ and $k_y$ axis will open to same direction, which is solely determined by the sign of  $t_{3}+t_{4}$.
When $\left|t_{3}+t_{4}\right|<\left|t_{3}-t_{4}\right|$, the two points are conventional  charge-2 Weyl points with type-I dispersion, and  the case of $\left|t_{3}+t_{4}\right|=\left|t_{3}-t_{4}\right|$ represents a critical state, which however generally is unstable.
It is worth mentioning that the hopping parameters $t_{i}$ ($i$ = 1,2,3,4) can be flexibly tuned  in sonic crystal. Thus, a sonic crystal is an ideal acoustic platform to experimentally realize the type-III Weyl phase.

Moreover, the chirality of the two Weyl points can be adjusted independently by tuning the model parameters $t_{1(2)}$.
Specifically, the two Weyl points have opposite chirality when  $|t_1|>|t_2|$, while have same chirality when  $|t_1|<|t_2|$. Therefore, for the case of  $|t_1|>|t_2|$, the two Weyl points at M and A points can be the only  band crossings of the system, leading to a single pair of Weyl points in nonmagnetic crystals.

\textit{\textcolor{blue}{Sonic crystal and single pair of type-III Weyl points}}\textit{.}
The sonic crystal sample is fabricated by 3D printing, as shown in Fig.~\ref{fig2}(a). The top and side views of the sample with a red outline of the unit cell are shown in Fig.~\ref{fig2}(b). The unit cell with lattice constant $a$= 30 mm and $H$= 30 mm is schematized in Fig.~\ref{fig2}(c), which consists of two pillars of diameter $D$=16 mm and height $h$=10 mm. Four tubes connect these pillars with radius $r_{1}$=2 mm, $r_{2}$=3 mm, $r_{3}$=2 mm, and $r_{4}$=3.5 mm, respectively. In addition, the corresponding bulk BZ (along with the two type-III Weyl points) is illustrated in Fig.~\ref{fig2}(d). The structure can be regarded as acoustically rigid and filled with air in the cavities, and can be used to simulate the novel phenomena in the spinless systems with the same crystalline symmetry.

We first simulate the band structure of the sample  to demonstrate  the existence of the type-III Weyl points. The result is presented in  Fig.~\ref{fig2}(e). We can clearly observe two Weyl points at $M$ and $A$, termed as $W_{M}$ and $W_{A}$, respectively.
$W_{A}$ and $W_{M}$ consist of two topological bands in the frequency range of 0 to 5.3 kHz without distractions from other trivial bands.
Notice that the two Weyl points are always pinned at the high symmetry points $M$ and $A$ due to $D_{4}$ symmetry, which in turn makes these points more easily observed in experiments.
Moreover, one can  find that $W_{A}$ and $W_{M}$ host linear dispersion along $k_{z}$ direction ($M$-$A$ path) and quadratic dispersion in $k_{x}$-$k_{y}$ plane. Hence, $W_{A}$ and $W_{M}$ are two charge-2 Weyl points, consistent with the symmetry analysis  and the  TB model discussed above.
The Chern number of $W_{A}$ and $W_{M}$ are calculated as ${\cal{C}}=-2$ and  ${\cal{C}}=2$ [see Fig.~\ref{fig2}(e)].

Particularly,  as  shown in Fig.~\ref{fig2}(e), parabolic dispersion of the conduction band and valence band open the same orientation along the $k_{x}$ ($k_{y}$) direction, therefore  $W_{A}$ and $W_{M}$ are the type-III Weyl points. In order to display the type-III dispersion more vividly, we plot the three-dimensional band structure of the $k_{x}$-$k_{y}$ plane around $M$ and $A$ points, as shown in Fig.~\ref{fig2}(f) and Fig.~\ref{fig2}(g).
Remarkably,  a careful scanning shows $W_{M}$ and $W_{A}$ are the only two degeneracies in the bulk BZ.
Thus, an ideal single pair of type-III Weyl points state has been presented.

\begin{figure}[t]
\includegraphics[width=7.0cm]{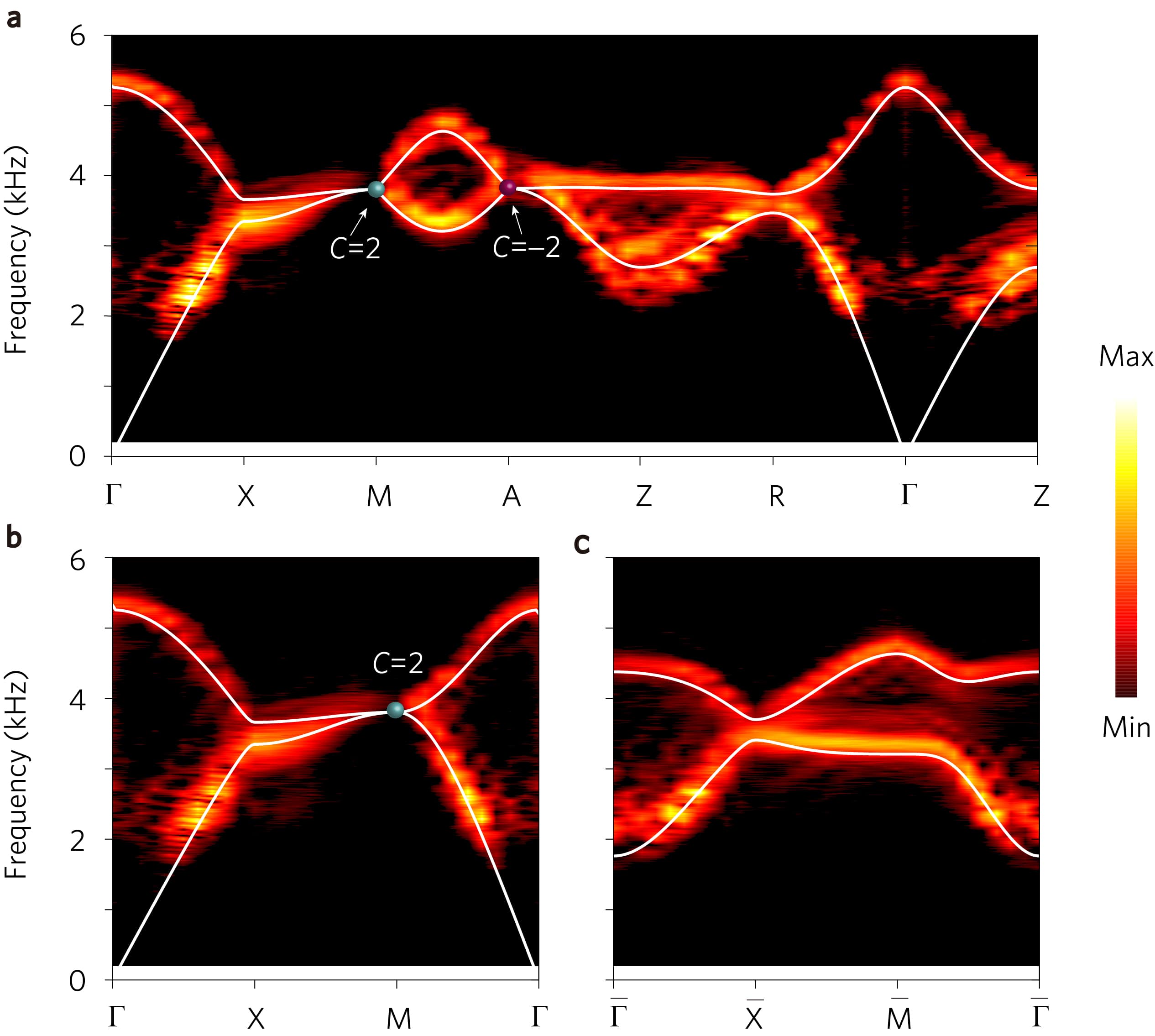} \caption{Experimental observation of the single pair  of the type-III Weyl points in the sonic crystal. (a) The measured and simulated bulk band structure along the high symmetry lines in the BZ.  (b) The bulk band structure along the path in $k_{z} = 0$ plane and the band structure of (c) in $k_{z} = 0.5\pi/H$ plane.
The color maps denote the magnitude of the experimental Fourier spectrum, and the white curves represent the simulated results.
\label{fig3}}
\end{figure}

\begin{figure}[t]
\includegraphics[width=8.0cm]{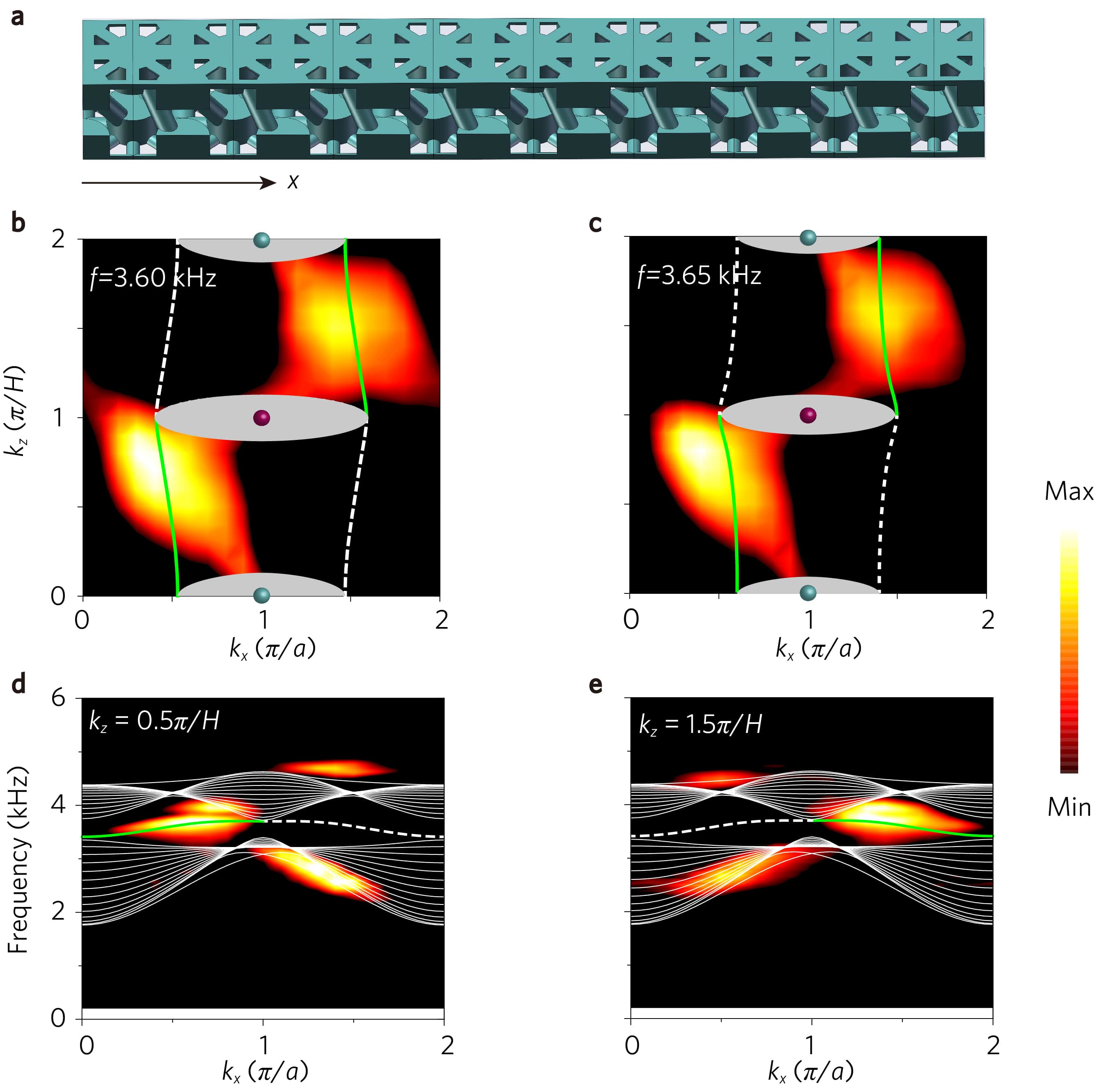} \caption{Topological acoustic Fermi arcs and surface states. (a) Schematic of the supercell for full-wave simulations of surface-state dispersions and acoustic Fermi arcs. (b)-(c) The isofrequency contours in the (100) surface BZ at $f$ = 3.60 kHz and $f$ = 3.65 kHz. The colors denote Fermi arc states that connect projected Weyl points and green solid lines represent the simulated results (the dashed lines represent opposite surfaces). (d)-(e) The Fourier transforms of the surface acoustic wave fields on the (001) surface with $k_{z}$ set as $0.5\pi/H$ and $1.5\pi/H$, respectively. The projected bulk bands are denoted by grey lines, and the surface states are displayed by the green solid lines.
\label{fig4}}
\end{figure}

To confirm the fact of the ideal type-III Weyl point  state in the  sonic crystal, we conduct experiments to obtain the dispersion curves. The acoustic source is placed at the center of the sample with $16\times16\times16$ unit cells. Using a tiny microphone attached to a thin steel rod, we can obtain the 3D acoustic fields of the sample at various frequencies (from 0.2 kHz to 8 kHz, with a step of 9.75 Hz). The color maps in Fig.~\ref{fig3} represent the experimental dispersions by 3D Fourier transforming the measured pressure fields inside the sample into the momentum space. We can find the measured results agree well with the computational simulation (denoted by white lines), two Weyl points  ($\mathcal{C}=\pm2$) locate at $M$ and $A$ can be observed at a frequency about 3.82 kHz [see Fig.~\ref{fig3}(a)]. More importantly, the dispersion characteristics of type-III 
also is clearly observed in Fig.~\ref{fig3}(b). Thus, the existence of a single pair of  type-III Weyl points in the sample is verified (more evidences can be found in Fig. \ref{fig4}). In addition, as depicted in Fig.~\ref{fig3}(c), we  measure the band structure in the plane $\overline{\Gamma}-\overline{X}-\overline{M}$ (with $k_{z}=0.5\pi/H$) that is far away from the type-III Weyl points. The result shows that the band crossing point will open a band gap and the slice at $k_{z}=0.5\pi/H$ therefore is a Chern insulator with  $|{\cal{C}}|=1$.

\textit{\textcolor{blue}{Fermi arcs and chiral edge states}}\textit{. }
An important feature of topological materials is the bulk-edge correspondence \cite{hatsugai1993chern,ryu2002topological}, that is, surface states can emerge at boundaries owning to the nontrivial topological properties of the bulk states. Because of the non-zero topological charge, type-III Weyl points are expected to present surface states similar to other Weyl semimetals. We then detect the topological surface wave of the sample. Specifically, a loudspeaker is placed at the center of the (100) surface, and we measure the acoustic field profile on that surface. The supercell in Fig.~\ref{fig4}(a) has a finite dimension along the $x$-direction but is periodically arranged along the $y$-direction and the $z$-direction. The boundary condition is composed of half unit cells. The experimentally measured surface Fermi arcs on isofrequency planes are shown in Fig.~\ref{fig4}(b)-(c) with frequencies $f$ = 3.60 kHz and $f$ = 3.65 kHz, respectively. The hot color map denotes the observed values, while white dashed curves (right boundary), green curves (left border), and gray areas show the predicted projected dispersions.
Because the fact of type-III Weyl points in the system possesses a topological charge of $\pm2$, one can observe that the two branches of Fermi arcs originating from a projected Weyl point link to another Weyl point with opposite chirality.
The appearance of only two Fermi arcs further demonstrates the existence of single pain of type-III Weyl point state.
Moreover, Fig.~\ref{fig4}(b)-(c) show the two Fermi arcs  forms a closed non-contractible loop, which is very extended and winds around the surface BZ along $k_z$ direction, in sharp contrast to the open Fermi arc in conventional (charge-1) Weyl point states.
%

Since the three-dimensional system can be regarded as a collection of two-dimensional subsystems at different  $k_{z}$,  we further plot the surface band structure of the sample at fixed $k_{z}$.  The results are comparable to the chiral edge states, as seen in Fig.~\ref{fig4}(d)-(e). When we fix $k_{z}$ as $0.5\pi/H$, the subsystems are equivalent two-dimensional sonic Chern insulators. We observe the chiral edge states connecting different sets of frequencies like the Chern insulator, as shown in Fig.~\ref{fig4}(d). Furthermore, we move the $k_{z}$ plane to 1.5 $\pi/H$, the Chern number of the subsystem of the two-dimensional Chern insulator has an opposite sign, as observed in Fig.~\ref{fig4}(e). Furthermore, the experimental measurements of the surface states also show good consistency with the numerical results.

\textit{\textcolor{blue}{Conclusion}}\textit{. }
In summary, we have designed and fabricated the first type-III acoustic Weyl sonic crystal, which contains only a pair of type-III Weyl points carrying a topological charge of $\pm2$ at the high symmetry points in the BZ. We observe two remarkable long Fermi arc states of type-III Weyl point in experiments and capture the chiral edge state of the two-dimensional Chern insulator at fixed $k_{z}$. With consistent simulated and experimental results, we have demonstrated the existence and dispersion characteristics of type-III Weyl points. 
Our system not only opens the door to experimentally observing the type-III Weyl points but also provides a viable platform for further exploring the physics associated with type-III Weyl points. 
The type-III Weyl points discussed in our acoustic semimetal can also be realized in photonic, mechanical and cold atom systems as well.

\acknowledgements
The work is supported by the National Key R\&D Program of China (Grant No. 2020YFA0308800), the National Natural Science Foundation of China (Grant Nos. 11734003, 12061131002, 12102039, 12004035), the Strategic Priority Research Program of Chinese Academy of Sciences (Grant No. XDB30000000) and the China Postdoctoral Science Foundation (Grant Nos.2020M680011, 2021T140057).
\bibliography{ref}

\begin{thebibliography}{51}%
\makeatletter
\providecommand \@ifxundefined [1]{%
 \@ifx{#1\undefined}
}%
\providecommand \@ifnum [1]{%
 \ifnum #1\expandafter \@firstoftwo
 \else \expandafter \@secondoftwo
 \fi
}%
\providecommand \@ifx [1]{%
 \ifx #1\expandafter \@firstoftwo
 \else \expandafter \@secondoftwo
 \fi
}%
\providecommand \natexlab [1]{#1}%
\providecommand \enquote  [1]{``#1''}%
\providecommand \bibnamefont  [1]{#1}%
\providecommand \bibfnamefont [1]{#1}%
\providecommand \citenamefont [1]{#1}%
\providecommand \href@noop [0]{\@secondoftwo}%
\providecommand \href [0]{\begingroup \@sanitize@url \@href}%
\providecommand \@href[1]{\@@startlink{#1}\@@href}%
\providecommand \@@href[1]{\endgroup#1\@@endlink}%
\providecommand \@sanitize@url [0]{\catcode `\\12\catcode `\$12\catcode
  `\&12\catcode `\#12\catcode `\^12\catcode `\_12\catcode `\%12\relax}%
\providecommand \@@startlink[1]{}%
\providecommand \@@endlink[0]{}%
\providecommand \url  [0]{\begingroup\@sanitize@url \@url }%
\providecommand \@url [1]{\endgroup\@href {#1}{\urlprefix }}%
\providecommand \urlprefix  [0]{URL }%
\providecommand \Eprint [0]{\href }%
\providecommand \doibase [0]{https://doi.org/}%
\providecommand \selectlanguage [0]{\@gobble}%
\providecommand \bibinfo  [0]{\@secondoftwo}%
\providecommand \bibfield  [0]{\@secondoftwo}%
\providecommand \translation [1]{[#1]}%
\providecommand \BibitemOpen [0]{}%
\providecommand \bibitemStop [0]{}%
\providecommand \bibitemNoStop [0]{.\EOS\space}%
\providecommand \EOS [0]{\spacefactor3000\relax}%
\providecommand \BibitemShut  [1]{\csname bibitem#1\endcsname}%
\let\auto@bib@innerbib\@empty
\bibitem [{\citenamefont {Armitage}\ \emph {et~al.}(2018)\citenamefont
  {Armitage}, \citenamefont {Mele},\ and\ \citenamefont
  {Vishwanath}}]{weylDirac}%
  \BibitemOpen
  \bibfield  {author} {\bibinfo {author} {\bibfnamefont {N.~P.}\ \bibnamefont
  {Armitage}}, \bibinfo {author} {\bibfnamefont {E.~J.}\ \bibnamefont {Mele}},\
  and\ \bibinfo {author} {\bibfnamefont {A.}~\bibnamefont {Vishwanath}},\
  }\bibfield  {title} {\bibinfo {title} {Weyl and dirac semimetals in
  three-dimensional solids},\ }\href@noop {} {\bibfield  {journal} {\bibinfo
  {journal} {Rev. Mod. Phys.}\ }\textbf {\bibinfo {volume} {90}},\ \bibinfo
  {pages} {015001} (\bibinfo {year} {2018})}\BibitemShut {NoStop}%
\bibitem [{\citenamefont {Murakami}(2007)}]{murakami2007phase}%
  \BibitemOpen
  \bibfield  {author} {\bibinfo {author} {\bibfnamefont {S.}~\bibnamefont
  {Murakami}},\ }\bibfield  {title} {\bibinfo {title} {Phase transition between
  the quantum spin hall and insulator phases in 3d: emergence of a topological
  gapless phase},\ }\href@noop {} {\bibfield  {journal} {\bibinfo  {journal}
  {New Journal of Physics}\ }\textbf {\bibinfo {volume} {9}},\ \bibinfo {pages}
  {356} (\bibinfo {year} {2007})}\BibitemShut {NoStop}%
\bibitem [{\citenamefont {Wan}\ \emph {et~al.}(2011)\citenamefont {Wan},
  \citenamefont {Turner}, \citenamefont {Vishwanath},\ and\ \citenamefont
  {Savrasov}}]{wan2011topological}%
  \BibitemOpen
  \bibfield  {author} {\bibinfo {author} {\bibfnamefont {X.}~\bibnamefont
  {Wan}}, \bibinfo {author} {\bibfnamefont {A.~M.}\ \bibnamefont {Turner}},
  \bibinfo {author} {\bibfnamefont {A.}~\bibnamefont {Vishwanath}},\ and\
  \bibinfo {author} {\bibfnamefont {S.~Y.}\ \bibnamefont {Savrasov}},\
  }\bibfield  {title} {\bibinfo {title} {Topological semimetal and fermi-arc
  surface states in the electronic structure of pyrochlore iridates},\
  }\href@noop {} {\bibfield  {journal} {\bibinfo  {journal} {Physical Review
  B}\ }\textbf {\bibinfo {volume} {83}},\ \bibinfo {pages} {205101} (\bibinfo
  {year} {2011})}\BibitemShut {NoStop}%
\bibitem [{\citenamefont {Weng}\ \emph {et~al.}(2015)\citenamefont {Weng},
  \citenamefont {Fang}, \citenamefont {Fang}, \citenamefont {Bernevig},\ and\
  \citenamefont {Dai}}]{weng2015weyl}%
  \BibitemOpen
  \bibfield  {author} {\bibinfo {author} {\bibfnamefont {H.}~\bibnamefont
  {Weng}}, \bibinfo {author} {\bibfnamefont {C.}~\bibnamefont {Fang}}, \bibinfo
  {author} {\bibfnamefont {Z.}~\bibnamefont {Fang}}, \bibinfo {author}
  {\bibfnamefont {B.~A.}\ \bibnamefont {Bernevig}},\ and\ \bibinfo {author}
  {\bibfnamefont {X.}~\bibnamefont {Dai}},\ }\bibfield  {title} {\bibinfo
  {title} {Weyl semimetal phase in noncentrosymmetric transition-metal
  monophosphides},\ }\href@noop {} {\bibfield  {journal} {\bibinfo  {journal}
  {Physical Review X}\ }\textbf {\bibinfo {volume} {5}},\ \bibinfo {pages}
  {011029} (\bibinfo {year} {2015})}\BibitemShut {NoStop}%
\bibitem [{\citenamefont {Yu}\ \emph {et~al.}(2022)\citenamefont {Yu},
  \citenamefont {Zhang}, \citenamefont {Liu}, \citenamefont {Wu}, \citenamefont
  {Li}, \citenamefont {Zhang}, \citenamefont {Yang},\ and\ \citenamefont
  {Yao}}]{yu2022encyclopedia}%
  \BibitemOpen
  \bibfield  {author} {\bibinfo {author} {\bibfnamefont {Z.-M.}\ \bibnamefont
  {Yu}}, \bibinfo {author} {\bibfnamefont {Z.}~\bibnamefont {Zhang}}, \bibinfo
  {author} {\bibfnamefont {G.-B.}\ \bibnamefont {Liu}}, \bibinfo {author}
  {\bibfnamefont {W.}~\bibnamefont {Wu}}, \bibinfo {author} {\bibfnamefont
  {X.-P.}\ \bibnamefont {Li}}, \bibinfo {author} {\bibfnamefont {R.-W.}\
  \bibnamefont {Zhang}}, \bibinfo {author} {\bibfnamefont {S.~A.}\ \bibnamefont
  {Yang}},\ and\ \bibinfo {author} {\bibfnamefont {Y.}~\bibnamefont {Yao}},\
  }\bibfield  {title} {\bibinfo {title} {Encyclopedia of emergent particles in
  three-dimensional crystals},\ }\href@noop {} {\bibfield  {journal} {\bibinfo
  {journal} {Science Bulletin}\ }\textbf {\bibinfo {volume} {67}},\ \bibinfo
  {pages} {375} (\bibinfo {year} {2022})}\BibitemShut {NoStop}%
\bibitem [{\citenamefont {Burkov}(2015)}]{burkov2015chiral}%
  \BibitemOpen
  \bibfield  {author} {\bibinfo {author} {\bibfnamefont {A.}~\bibnamefont
  {Burkov}},\ }\bibfield  {title} {\bibinfo {title} {Chiral anomaly and
  transport in weyl metals},\ }\href@noop {} {\bibfield  {journal} {\bibinfo
  {journal} {Journal of Physics: Condensed Matter}\ }\textbf {\bibinfo {volume}
  {27}},\ \bibinfo {pages} {113201} (\bibinfo {year} {2015})}\BibitemShut
  {NoStop}%
\bibitem [{\citenamefont {Liu}\ \emph {et~al.}(2020)\citenamefont {Liu},
  \citenamefont {Yu}, \citenamefont {Xiao},\ and\ \citenamefont
  {Yang}}]{liu2020quantized}%
  \BibitemOpen
  \bibfield  {author} {\bibinfo {author} {\bibfnamefont {Y.}~\bibnamefont
  {Liu}}, \bibinfo {author} {\bibfnamefont {Z.-M.}\ \bibnamefont {Yu}},
  \bibinfo {author} {\bibfnamefont {C.}~\bibnamefont {Xiao}},\ and\ \bibinfo
  {author} {\bibfnamefont {S.~A.}\ \bibnamefont {Yang}},\ }\bibfield  {title}
  {\bibinfo {title} {Quantized circulation of anomalous shift in interface
  reflection},\ }\href@noop {} {\bibfield  {journal} {\bibinfo  {journal}
  {Physical Review Letters}\ }\textbf {\bibinfo {volume} {125}},\ \bibinfo
  {pages} {076801} (\bibinfo {year} {2020})}\BibitemShut {NoStop}%
\bibitem [{\citenamefont {Nagaosa}\ \emph {et~al.}(2020)\citenamefont
  {Nagaosa}, \citenamefont {Morimoto},\ and\ \citenamefont
  {Tokura}}]{nagaosa2020transport}%
  \BibitemOpen
  \bibfield  {author} {\bibinfo {author} {\bibfnamefont {N.}~\bibnamefont
  {Nagaosa}}, \bibinfo {author} {\bibfnamefont {T.}~\bibnamefont {Morimoto}},\
  and\ \bibinfo {author} {\bibfnamefont {Y.}~\bibnamefont {Tokura}},\
  }\bibfield  {title} {\bibinfo {title} {Transport, magnetic and optical
  properties of weyl materials},\ }\href@noop {} {\bibfield  {journal}
  {\bibinfo  {journal} {Nature Reviews Materials}\ }\textbf {\bibinfo {volume}
  {5}},\ \bibinfo {pages} {621} (\bibinfo {year} {2020})}\BibitemShut {NoStop}%
\bibitem [{\citenamefont {de~Juan}\ \emph {et~al.}(2017)\citenamefont
  {de~Juan}, \citenamefont {Grushin}, \citenamefont {Morimoto},\ and\
  \citenamefont {Moore}}]{de2017quantized}%
  \BibitemOpen
  \bibfield  {author} {\bibinfo {author} {\bibfnamefont {F.}~\bibnamefont
  {de~Juan}}, \bibinfo {author} {\bibfnamefont {A.~G.}\ \bibnamefont
  {Grushin}}, \bibinfo {author} {\bibfnamefont {T.}~\bibnamefont {Morimoto}},\
  and\ \bibinfo {author} {\bibfnamefont {J.~E.}\ \bibnamefont {Moore}},\
  }\bibfield  {title} {\bibinfo {title} {Quantized circular photogalvanic
  effect in weyl semimetals},\ }\href@noop {} {\bibfield  {journal} {\bibinfo
  {journal} {Nature communications}\ }\textbf {\bibinfo {volume} {8}},\
  \bibinfo {pages} {1} (\bibinfo {year} {2017})}\BibitemShut {NoStop}%
\bibitem [{\citenamefont {Zhou}\ \emph {et~al.}(2015)\citenamefont {Zhou},
  \citenamefont {Chang},\ and\ \citenamefont {Xiao}}]{zhou2015plasmon}%
  \BibitemOpen
  \bibfield  {author} {\bibinfo {author} {\bibfnamefont {J.}~\bibnamefont
  {Zhou}}, \bibinfo {author} {\bibfnamefont {H.-R.}\ \bibnamefont {Chang}},\
  and\ \bibinfo {author} {\bibfnamefont {D.}~\bibnamefont {Xiao}},\ }\bibfield
  {title} {\bibinfo {title} {Plasmon mode as a detection of the chiral anomaly
  in weyl semimetals},\ }\href@noop {} {\bibfield  {journal} {\bibinfo
  {journal} {Physical Review B}\ }\textbf {\bibinfo {volume} {91}},\ \bibinfo
  {pages} {035114} (\bibinfo {year} {2015})}\BibitemShut {NoStop}%
\bibitem [{\citenamefont {Fang}\ \emph {et~al.}(2012)\citenamefont {Fang},
  \citenamefont {Gilbert}, \citenamefont {Dai},\ and\ \citenamefont
  {Bernevig}}]{fang2012multi}%
  \BibitemOpen
  \bibfield  {author} {\bibinfo {author} {\bibfnamefont {C.}~\bibnamefont
  {Fang}}, \bibinfo {author} {\bibfnamefont {M.~J.}\ \bibnamefont {Gilbert}},
  \bibinfo {author} {\bibfnamefont {X.}~\bibnamefont {Dai}},\ and\ \bibinfo
  {author} {\bibfnamefont {B.~A.}\ \bibnamefont {Bernevig}},\ }\bibfield
  {title} {\bibinfo {title} {Multi-weyl topological semimetals stabilized by
  point group symmetry},\ }\href@noop {} {\bibfield  {journal} {\bibinfo
  {journal} {Physical review letters}\ }\textbf {\bibinfo {volume} {108}},\
  \bibinfo {pages} {266802} (\bibinfo {year} {2012})}\BibitemShut {NoStop}%
\bibitem [{\citenamefont {Zhang}\ \emph {et~al.}(2020)\citenamefont {Zhang},
  \citenamefont {Takahashi}, \citenamefont {Fang},\ and\ \citenamefont
  {Murakami}}]{zhang2020twofold}%
  \BibitemOpen
  \bibfield  {author} {\bibinfo {author} {\bibfnamefont {T.}~\bibnamefont
  {Zhang}}, \bibinfo {author} {\bibfnamefont {R.}~\bibnamefont {Takahashi}},
  \bibinfo {author} {\bibfnamefont {C.}~\bibnamefont {Fang}},\ and\ \bibinfo
  {author} {\bibfnamefont {S.}~\bibnamefont {Murakami}},\ }\bibfield  {title}
  {\bibinfo {title} {Twofold quadruple weyl nodes in chiral cubic crystals},\
  }\href@noop {} {\bibfield  {journal} {\bibinfo  {journal} {Physical Review
  B}\ }\textbf {\bibinfo {volume} {102}},\ \bibinfo {pages} {125148} (\bibinfo
  {year} {2020})}\BibitemShut {NoStop}%
\bibitem [{\citenamefont {Cui}\ \emph {et~al.}(2021)\citenamefont {Cui},
  \citenamefont {Li}, \citenamefont {Ma}, \citenamefont {Yu},\ and\
  \citenamefont {Yao}}]{cui2021charge}%
  \BibitemOpen
  \bibfield  {author} {\bibinfo {author} {\bibfnamefont {C.}~\bibnamefont
  {Cui}}, \bibinfo {author} {\bibfnamefont {X.-P.}\ \bibnamefont {Li}},
  \bibinfo {author} {\bibfnamefont {D.-S.}\ \bibnamefont {Ma}}, \bibinfo
  {author} {\bibfnamefont {Z.-M.}\ \bibnamefont {Yu}},\ and\ \bibinfo {author}
  {\bibfnamefont {Y.}~\bibnamefont {Yao}},\ }\bibfield  {title} {\bibinfo
  {title} {Charge-four weyl point: Minimum lattice model and
  chirality-dependent properties},\ }\href@noop {} {\bibfield  {journal}
  {\bibinfo  {journal} {Physical Review B}\ }\textbf {\bibinfo {volume}
  {104}},\ \bibinfo {pages} {075115} (\bibinfo {year} {2021})}\BibitemShut
  {NoStop}%
\bibitem [{\citenamefont {Han}\ \emph {et~al.}(2019)\citenamefont {Han},
  \citenamefont {Lee}, \citenamefont {Moon},\ and\ \citenamefont
  {Min}}]{han2019emergent}%
  \BibitemOpen
  \bibfield  {author} {\bibinfo {author} {\bibfnamefont {S.}~\bibnamefont
  {Han}}, \bibinfo {author} {\bibfnamefont {C.}~\bibnamefont {Lee}}, \bibinfo
  {author} {\bibfnamefont {E.-G.}\ \bibnamefont {Moon}},\ and\ \bibinfo
  {author} {\bibfnamefont {H.}~\bibnamefont {Min}},\ }\bibfield  {title}
  {\bibinfo {title} {Emergent anisotropic non-fermi liquid at a topological
  phase transition in three dimensions},\ }\href@noop {} {\bibfield  {journal}
  {\bibinfo  {journal} {Physical review letters}\ }\textbf {\bibinfo {volume}
  {122}},\ \bibinfo {pages} {187601} (\bibinfo {year} {2019})}\BibitemShut
  {NoStop}%
\bibitem [{\citenamefont {Wang}\ \emph {et~al.}(2019)\citenamefont {Wang},
  \citenamefont {Liu},\ and\ \citenamefont {Zhang}}]{wang2019topological}%
  \BibitemOpen
  \bibfield  {author} {\bibinfo {author} {\bibfnamefont {J.-R.}\ \bibnamefont
  {Wang}}, \bibinfo {author} {\bibfnamefont {G.-Z.}\ \bibnamefont {Liu}},\ and\
  \bibinfo {author} {\bibfnamefont {C.-J.}\ \bibnamefont {Zhang}},\ }\bibfield
  {title} {\bibinfo {title} {Topological quantum critical point in a
  triple-weyl semimetal: Non-fermi-liquid behavior and instabilities},\
  }\href@noop {} {\bibfield  {journal} {\bibinfo  {journal} {Physical Review
  B}\ }\textbf {\bibinfo {volume} {99}},\ \bibinfo {pages} {195119} (\bibinfo
  {year} {2019})}\BibitemShut {NoStop}%
\bibitem [{\citenamefont {Zhang}\ \emph {et~al.}(2021)\citenamefont {Zhang},
  \citenamefont {Jian},\ and\ \citenamefont {Yao}}]{zhang2021quantum}%
  \BibitemOpen
  \bibfield  {author} {\bibinfo {author} {\bibfnamefont {S.-X.}\ \bibnamefont
  {Zhang}}, \bibinfo {author} {\bibfnamefont {S.-K.}\ \bibnamefont {Jian}},\
  and\ \bibinfo {author} {\bibfnamefont {H.}~\bibnamefont {Yao}},\ }\bibfield
  {title} {\bibinfo {title} {Quantum criticality preempted by nematicity},\
  }\href@noop {} {\bibfield  {journal} {\bibinfo  {journal} {Physical Review
  B}\ }\textbf {\bibinfo {volume} {103}},\ \bibinfo {pages} {165129} (\bibinfo
  {year} {2021})}\BibitemShut {NoStop}%
\bibitem [{\citenamefont {Soluyanov}\ \emph {et~al.}(2015)\citenamefont
  {Soluyanov}, \citenamefont {Gresch}, \citenamefont {Wang}, \citenamefont
  {Wu}, \citenamefont {Troyer}, \citenamefont {Dai},\ and\ \citenamefont
  {Bernevig}}]{soluyanov2015type}%
  \BibitemOpen
  \bibfield  {author} {\bibinfo {author} {\bibfnamefont {A.~A.}\ \bibnamefont
  {Soluyanov}}, \bibinfo {author} {\bibfnamefont {D.}~\bibnamefont {Gresch}},
  \bibinfo {author} {\bibfnamefont {Z.}~\bibnamefont {Wang}}, \bibinfo {author}
  {\bibfnamefont {Q.}~\bibnamefont {Wu}}, \bibinfo {author} {\bibfnamefont
  {M.}~\bibnamefont {Troyer}}, \bibinfo {author} {\bibfnamefont
  {X.}~\bibnamefont {Dai}},\ and\ \bibinfo {author} {\bibfnamefont {B.~A.}\
  \bibnamefont {Bernevig}},\ }\bibfield  {title} {\bibinfo {title} {Type-ii
  weyl semimetals},\ }\href@noop {} {\bibfield  {journal} {\bibinfo  {journal}
  {Nature}\ }\textbf {\bibinfo {volume} {527}},\ \bibinfo {pages} {495}
  (\bibinfo {year} {2015})}\BibitemShut {NoStop}%
\bibitem [{\citenamefont {Li}\ \emph {et~al.}(2021{\natexlab{a}})\citenamefont
  {Li}, \citenamefont {Deng}, \citenamefont {Fu}, \citenamefont {Li},
  \citenamefont {Ma}, \citenamefont {Han}, \citenamefont {Zhou}, \citenamefont
  {Zhou},\ and\ \citenamefont {Yao}}]{li2021type}%
  \BibitemOpen
  \bibfield  {author} {\bibinfo {author} {\bibfnamefont {X.-P.}\ \bibnamefont
  {Li}}, \bibinfo {author} {\bibfnamefont {K.}~\bibnamefont {Deng}}, \bibinfo
  {author} {\bibfnamefont {B.}~\bibnamefont {Fu}}, \bibinfo {author}
  {\bibfnamefont {Y.}~\bibnamefont {Li}}, \bibinfo {author} {\bibfnamefont
  {D.-S.}\ \bibnamefont {Ma}}, \bibinfo {author} {\bibfnamefont
  {J.}~\bibnamefont {Han}}, \bibinfo {author} {\bibfnamefont {J.}~\bibnamefont
  {Zhou}}, \bibinfo {author} {\bibfnamefont {S.}~\bibnamefont {Zhou}},\ and\
  \bibinfo {author} {\bibfnamefont {Y.}~\bibnamefont {Yao}},\ }\bibfield
  {title} {\bibinfo {title} {Type-iii weyl semimetals:(tase 4) 2 i},\
  }\href@noop {} {\bibfield  {journal} {\bibinfo  {journal} {Physical Review
  B}\ }\textbf {\bibinfo {volume} {103}},\ \bibinfo {pages} {L081402} (\bibinfo
  {year} {2021}{\natexlab{a}})}\BibitemShut {NoStop}%
\bibitem [{\citenamefont {O’Brien}\ \emph {et~al.}(2016)\citenamefont
  {O’Brien}, \citenamefont {Diez},\ and\ \citenamefont
  {Beenakker}}]{o2016magnetic}%
  \BibitemOpen
  \bibfield  {author} {\bibinfo {author} {\bibfnamefont {T.}~\bibnamefont
  {O’Brien}}, \bibinfo {author} {\bibfnamefont {M.}~\bibnamefont {Diez}},\
  and\ \bibinfo {author} {\bibfnamefont {C.}~\bibnamefont {Beenakker}},\
  }\bibfield  {title} {\bibinfo {title} {Magnetic breakdown and klein tunneling
  in a type-ii weyl semimetal},\ }\href@noop {} {\bibfield  {journal} {\bibinfo
   {journal} {Physical review letters}\ }\textbf {\bibinfo {volume} {116}},\
  \bibinfo {pages} {236401} (\bibinfo {year} {2016})}\BibitemShut {NoStop}%
\bibitem [{\citenamefont {Yu}\ \emph {et~al.}(2016)\citenamefont {Yu},
  \citenamefont {Yao},\ and\ \citenamefont {Yang}}]{yu2016predicted}%
  \BibitemOpen
  \bibfield  {author} {\bibinfo {author} {\bibfnamefont {Z.-M.}\ \bibnamefont
  {Yu}}, \bibinfo {author} {\bibfnamefont {Y.}~\bibnamefont {Yao}},\ and\
  \bibinfo {author} {\bibfnamefont {S.~A.}\ \bibnamefont {Yang}},\ }\bibfield
  {title} {\bibinfo {title} {Predicted unusual magnetoresponse in type-ii weyl
  semimetals},\ }\href@noop {} {\bibfield  {journal} {\bibinfo  {journal}
  {Physical review letters}\ }\textbf {\bibinfo {volume} {117}},\ \bibinfo
  {pages} {077202} (\bibinfo {year} {2016})}\BibitemShut {NoStop}%
\bibitem [{\citenamefont {Tchoumakov}\ \emph {et~al.}(2016)\citenamefont
  {Tchoumakov}, \citenamefont {Civelli},\ and\ \citenamefont
  {Goerbig}}]{tchoumakov2016magnetic}%
  \BibitemOpen
  \bibfield  {author} {\bibinfo {author} {\bibfnamefont {S.}~\bibnamefont
  {Tchoumakov}}, \bibinfo {author} {\bibfnamefont {M.}~\bibnamefont
  {Civelli}},\ and\ \bibinfo {author} {\bibfnamefont {M.~O.}\ \bibnamefont
  {Goerbig}},\ }\bibfield  {title} {\bibinfo {title} {Magnetic-field-induced
  relativistic properties in type-i and type-ii weyl semimetals},\ }\href@noop
  {} {\bibfield  {journal} {\bibinfo  {journal} {Physical review letters}\
  }\textbf {\bibinfo {volume} {117}},\ \bibinfo {pages} {086402} (\bibinfo
  {year} {2016})}\BibitemShut {NoStop}%
\bibitem [{\citenamefont {Udagawa}\ and\ \citenamefont
  {Bergholtz}(2016)}]{udagawa2016field}%
  \BibitemOpen
  \bibfield  {author} {\bibinfo {author} {\bibfnamefont {M.}~\bibnamefont
  {Udagawa}}\ and\ \bibinfo {author} {\bibfnamefont {E.~J.}\ \bibnamefont
  {Bergholtz}},\ }\bibfield  {title} {\bibinfo {title} {Field-selective anomaly
  and chiral mode reversal in type-ii weyl materials},\ }\href@noop {}
  {\bibfield  {journal} {\bibinfo  {journal} {Physical review letters}\
  }\textbf {\bibinfo {volume} {117}},\ \bibinfo {pages} {086401} (\bibinfo
  {year} {2016})}\BibitemShut {NoStop}%
\bibitem [{\citenamefont {Lv}\ \emph {et~al.}(2021)\citenamefont {Lv},
  \citenamefont {Qian},\ and\ \citenamefont {Ding}}]{lv2021experimental}%
  \BibitemOpen
  \bibfield  {author} {\bibinfo {author} {\bibfnamefont {B.}~\bibnamefont
  {Lv}}, \bibinfo {author} {\bibfnamefont {T.}~\bibnamefont {Qian}},\ and\
  \bibinfo {author} {\bibfnamefont {H.}~\bibnamefont {Ding}},\ }\bibfield
  {title} {\bibinfo {title} {Experimental perspective on three-dimensional
  topological semimetals},\ }\href@noop {} {\bibfield  {journal} {\bibinfo
  {journal} {Reviews of Modern Physics}\ }\textbf {\bibinfo {volume} {93}},\
  \bibinfo {pages} {025002} (\bibinfo {year} {2021})}\BibitemShut {NoStop}%
\bibitem [{\citenamefont {Lv}\ \emph {et~al.}(2015)\citenamefont {Lv},
  \citenamefont {Weng}, \citenamefont {Fu}, \citenamefont {Wang}, \citenamefont
  {Miao}, \citenamefont {Ma}, \citenamefont {Richard}, \citenamefont {Huang},
  \citenamefont {Zhao}, \citenamefont {Chen} \emph
  {et~al.}}]{lv2015experimental}%
  \BibitemOpen
  \bibfield  {author} {\bibinfo {author} {\bibfnamefont {B.}~\bibnamefont
  {Lv}}, \bibinfo {author} {\bibfnamefont {H.}~\bibnamefont {Weng}}, \bibinfo
  {author} {\bibfnamefont {B.}~\bibnamefont {Fu}}, \bibinfo {author}
  {\bibfnamefont {X.~P.}\ \bibnamefont {Wang}}, \bibinfo {author}
  {\bibfnamefont {H.}~\bibnamefont {Miao}}, \bibinfo {author} {\bibfnamefont
  {J.}~\bibnamefont {Ma}}, \bibinfo {author} {\bibfnamefont {P.}~\bibnamefont
  {Richard}}, \bibinfo {author} {\bibfnamefont {X.}~\bibnamefont {Huang}},
  \bibinfo {author} {\bibfnamefont {L.}~\bibnamefont {Zhao}}, \bibinfo {author}
  {\bibfnamefont {G.}~\bibnamefont {Chen}}, \emph {et~al.},\ }\bibfield
  {title} {\bibinfo {title} {Experimental discovery of weyl semimetal taas},\
  }\href@noop {} {\bibfield  {journal} {\bibinfo  {journal} {Physical Review
  X}\ }\textbf {\bibinfo {volume} {5}},\ \bibinfo {pages} {031013} (\bibinfo
  {year} {2015})}\BibitemShut {NoStop}%
\bibitem [{\citenamefont {Deng}\ \emph {et~al.}(2016)\citenamefont {Deng},
  \citenamefont {Wan}, \citenamefont {Deng}, \citenamefont {Zhang},
  \citenamefont {Ding}, \citenamefont {Wang}, \citenamefont {Yan},
  \citenamefont {Huang}, \citenamefont {Zhang}, \citenamefont {Xu} \emph
  {et~al.}}]{deng2016experimental}%
  \BibitemOpen
  \bibfield  {author} {\bibinfo {author} {\bibfnamefont {K.}~\bibnamefont
  {Deng}}, \bibinfo {author} {\bibfnamefont {G.}~\bibnamefont {Wan}}, \bibinfo
  {author} {\bibfnamefont {P.}~\bibnamefont {Deng}}, \bibinfo {author}
  {\bibfnamefont {K.}~\bibnamefont {Zhang}}, \bibinfo {author} {\bibfnamefont
  {S.}~\bibnamefont {Ding}}, \bibinfo {author} {\bibfnamefont {E.}~\bibnamefont
  {Wang}}, \bibinfo {author} {\bibfnamefont {M.}~\bibnamefont {Yan}}, \bibinfo
  {author} {\bibfnamefont {H.}~\bibnamefont {Huang}}, \bibinfo {author}
  {\bibfnamefont {H.}~\bibnamefont {Zhang}}, \bibinfo {author} {\bibfnamefont
  {Z.}~\bibnamefont {Xu}}, \emph {et~al.},\ }\bibfield  {title} {\bibinfo
  {title} {Experimental observation of topological fermi arcs in type-ii weyl
  semimetal mote2},\ }\href@noop {} {\bibfield  {journal} {\bibinfo  {journal}
  {Nature Physics}\ }\textbf {\bibinfo {volume} {12}},\ \bibinfo {pages} {1105}
  (\bibinfo {year} {2016})}\BibitemShut {NoStop}%
\bibitem [{\citenamefont {Huang}\ \emph
  {et~al.}(2016{\natexlab{a}})\citenamefont {Huang}, \citenamefont {McCormick},
  \citenamefont {Ochi}, \citenamefont {Zhao}, \citenamefont {Suzuki},
  \citenamefont {Arita}, \citenamefont {Wu}, \citenamefont {Mou}, \citenamefont
  {Cao}, \citenamefont {Yan} \emph {et~al.}}]{huang2016spectroscopic}%
  \BibitemOpen
  \bibfield  {author} {\bibinfo {author} {\bibfnamefont {L.}~\bibnamefont
  {Huang}}, \bibinfo {author} {\bibfnamefont {T.~M.}\ \bibnamefont
  {McCormick}}, \bibinfo {author} {\bibfnamefont {M.}~\bibnamefont {Ochi}},
  \bibinfo {author} {\bibfnamefont {Z.}~\bibnamefont {Zhao}}, \bibinfo {author}
  {\bibfnamefont {M.-T.}\ \bibnamefont {Suzuki}}, \bibinfo {author}
  {\bibfnamefont {R.}~\bibnamefont {Arita}}, \bibinfo {author} {\bibfnamefont
  {Y.}~\bibnamefont {Wu}}, \bibinfo {author} {\bibfnamefont {D.}~\bibnamefont
  {Mou}}, \bibinfo {author} {\bibfnamefont {H.}~\bibnamefont {Cao}}, \bibinfo
  {author} {\bibfnamefont {J.}~\bibnamefont {Yan}}, \emph {et~al.},\ }\bibfield
   {title} {\bibinfo {title} {Spectroscopic evidence for a type ii weyl
  semimetallic state in mote2},\ }\href@noop {} {\bibfield  {journal} {\bibinfo
   {journal} {Nature materials}\ }\textbf {\bibinfo {volume} {15}},\ \bibinfo
  {pages} {1155} (\bibinfo {year} {2016}{\natexlab{a}})}\BibitemShut {NoStop}%
\bibitem [{\citenamefont {Yao}\ \emph {et~al.}(2019)\citenamefont {Yao},
  \citenamefont {Xu}, \citenamefont {Wu}, \citenamefont {Aut{\`e}s},
  \citenamefont {Kumar}, \citenamefont {Strocov}, \citenamefont {Plumb},
  \citenamefont {Radovic}, \citenamefont {Yazyev}, \citenamefont {Felser} \emph
  {et~al.}}]{yao2019observation}%
  \BibitemOpen
  \bibfield  {author} {\bibinfo {author} {\bibfnamefont {M.-Y.}\ \bibnamefont
  {Yao}}, \bibinfo {author} {\bibfnamefont {N.}~\bibnamefont {Xu}}, \bibinfo
  {author} {\bibfnamefont {Q.}~\bibnamefont {Wu}}, \bibinfo {author}
  {\bibfnamefont {G.}~\bibnamefont {Aut{\`e}s}}, \bibinfo {author}
  {\bibfnamefont {N.}~\bibnamefont {Kumar}}, \bibinfo {author} {\bibfnamefont
  {V.~N.}\ \bibnamefont {Strocov}}, \bibinfo {author} {\bibfnamefont {N.~C.}\
  \bibnamefont {Plumb}}, \bibinfo {author} {\bibfnamefont {M.}~\bibnamefont
  {Radovic}}, \bibinfo {author} {\bibfnamefont {O.~V.}\ \bibnamefont {Yazyev}},
  \bibinfo {author} {\bibfnamefont {C.}~\bibnamefont {Felser}}, \emph
  {et~al.},\ }\bibfield  {title} {\bibinfo {title} {Observation of weyl nodes
  in robust type-ii weyl semimetal wp 2},\ }\href@noop {} {\bibfield  {journal}
  {\bibinfo  {journal} {Physical Review Letters}\ }\textbf {\bibinfo {volume}
  {122}},\ \bibinfo {pages} {176402} (\bibinfo {year} {2019})}\BibitemShut
  {NoStop}%
\bibitem [{\citenamefont {Xu}\ \emph {et~al.}(2017)\citenamefont {Xu},
  \citenamefont {Alidoust}, \citenamefont {Chang}, \citenamefont {Lu},
  \citenamefont {Singh}, \citenamefont {Belopolski}, \citenamefont {Sanchez},
  \citenamefont {Zhang}, \citenamefont {Bian}, \citenamefont {Zheng} \emph
  {et~al.}}]{xu2017discovery}%
  \BibitemOpen
  \bibfield  {author} {\bibinfo {author} {\bibfnamefont {S.-Y.}\ \bibnamefont
  {Xu}}, \bibinfo {author} {\bibfnamefont {N.}~\bibnamefont {Alidoust}},
  \bibinfo {author} {\bibfnamefont {G.}~\bibnamefont {Chang}}, \bibinfo
  {author} {\bibfnamefont {H.}~\bibnamefont {Lu}}, \bibinfo {author}
  {\bibfnamefont {B.}~\bibnamefont {Singh}}, \bibinfo {author} {\bibfnamefont
  {I.}~\bibnamefont {Belopolski}}, \bibinfo {author} {\bibfnamefont {D.~S.}\
  \bibnamefont {Sanchez}}, \bibinfo {author} {\bibfnamefont {X.}~\bibnamefont
  {Zhang}}, \bibinfo {author} {\bibfnamefont {G.}~\bibnamefont {Bian}},
  \bibinfo {author} {\bibfnamefont {H.}~\bibnamefont {Zheng}}, \emph {et~al.},\
  }\bibfield  {title} {\bibinfo {title} {Discovery of lorentz-violating type ii
  weyl fermions in laalge},\ }\href@noop {} {\bibfield  {journal} {\bibinfo
  {journal} {Science advances}\ }\textbf {\bibinfo {volume} {3}},\ \bibinfo
  {pages} {e1603266} (\bibinfo {year} {2017})}\BibitemShut {NoStop}%
\bibitem [{\citenamefont {Chang}\ \emph {et~al.}(2016)\citenamefont {Chang},
  \citenamefont {Xu}, \citenamefont {Sanchez}, \citenamefont {Huang},
  \citenamefont {Lee}, \citenamefont {Chang}, \citenamefont {Bian},
  \citenamefont {Zheng}, \citenamefont {Belopolski}, \citenamefont {Alidoust}
  \emph {et~al.}}]{chang2016strongly}%
  \BibitemOpen
  \bibfield  {author} {\bibinfo {author} {\bibfnamefont {G.}~\bibnamefont
  {Chang}}, \bibinfo {author} {\bibfnamefont {S.-Y.}\ \bibnamefont {Xu}},
  \bibinfo {author} {\bibfnamefont {D.~S.}\ \bibnamefont {Sanchez}}, \bibinfo
  {author} {\bibfnamefont {S.-M.}\ \bibnamefont {Huang}}, \bibinfo {author}
  {\bibfnamefont {C.-C.}\ \bibnamefont {Lee}}, \bibinfo {author} {\bibfnamefont
  {T.-R.}\ \bibnamefont {Chang}}, \bibinfo {author} {\bibfnamefont
  {G.}~\bibnamefont {Bian}}, \bibinfo {author} {\bibfnamefont {H.}~\bibnamefont
  {Zheng}}, \bibinfo {author} {\bibfnamefont {I.}~\bibnamefont {Belopolski}},
  \bibinfo {author} {\bibfnamefont {N.}~\bibnamefont {Alidoust}}, \emph
  {et~al.},\ }\bibfield  {title} {\bibinfo {title} {A strongly robust type ii
  weyl fermion semimetal state in ta3s2},\ }\href@noop {} {\bibfield  {journal}
  {\bibinfo  {journal} {Science Advances}\ }\textbf {\bibinfo {volume} {2}},\
  \bibinfo {pages} {e1600295} (\bibinfo {year} {2016})}\BibitemShut {NoStop}%
\bibitem [{\citenamefont {Belopolski}\ \emph {et~al.}(2017)\citenamefont
  {Belopolski}, \citenamefont {Yu}, \citenamefont {Sanchez}, \citenamefont
  {Ishida}, \citenamefont {Chang}, \citenamefont {Zhang}, \citenamefont {Xu},
  \citenamefont {Zheng}, \citenamefont {Chang}, \citenamefont {Bian} \emph
  {et~al.}}]{belopolski2017signatures}%
  \BibitemOpen
  \bibfield  {author} {\bibinfo {author} {\bibfnamefont {I.}~\bibnamefont
  {Belopolski}}, \bibinfo {author} {\bibfnamefont {P.}~\bibnamefont {Yu}},
  \bibinfo {author} {\bibfnamefont {D.~S.}\ \bibnamefont {Sanchez}}, \bibinfo
  {author} {\bibfnamefont {Y.}~\bibnamefont {Ishida}}, \bibinfo {author}
  {\bibfnamefont {T.-R.}\ \bibnamefont {Chang}}, \bibinfo {author}
  {\bibfnamefont {S.~S.}\ \bibnamefont {Zhang}}, \bibinfo {author}
  {\bibfnamefont {S.-Y.}\ \bibnamefont {Xu}}, \bibinfo {author} {\bibfnamefont
  {H.}~\bibnamefont {Zheng}}, \bibinfo {author} {\bibfnamefont
  {G.}~\bibnamefont {Chang}}, \bibinfo {author} {\bibfnamefont
  {G.}~\bibnamefont {Bian}}, \emph {et~al.},\ }\bibfield  {title} {\bibinfo
  {title} {Signatures of a time-reversal symmetric weyl semimetal with only
  four weyl points},\ }\href@noop {} {\bibfield  {journal} {\bibinfo  {journal}
  {Nature communications}\ }\textbf {\bibinfo {volume} {8}},\ \bibinfo {pages}
  {1} (\bibinfo {year} {2017})}\BibitemShut {NoStop}%
\bibitem [{\citenamefont {Huang}\ \emph
  {et~al.}(2016{\natexlab{b}})\citenamefont {Huang}, \citenamefont {Xu},
  \citenamefont {Belopolski}, \citenamefont {Lee}, \citenamefont {Chang},
  \citenamefont {Chang}, \citenamefont {Wang}, \citenamefont {Alidoust},
  \citenamefont {Bian}, \citenamefont {Neupane} \emph {et~al.}}]{huang2016new}%
  \BibitemOpen
  \bibfield  {author} {\bibinfo {author} {\bibfnamefont {S.-M.}\ \bibnamefont
  {Huang}}, \bibinfo {author} {\bibfnamefont {S.-Y.}\ \bibnamefont {Xu}},
  \bibinfo {author} {\bibfnamefont {I.}~\bibnamefont {Belopolski}}, \bibinfo
  {author} {\bibfnamefont {C.-C.}\ \bibnamefont {Lee}}, \bibinfo {author}
  {\bibfnamefont {G.}~\bibnamefont {Chang}}, \bibinfo {author} {\bibfnamefont
  {T.-R.}\ \bibnamefont {Chang}}, \bibinfo {author} {\bibfnamefont
  {B.}~\bibnamefont {Wang}}, \bibinfo {author} {\bibfnamefont {N.}~\bibnamefont
  {Alidoust}}, \bibinfo {author} {\bibfnamefont {G.}~\bibnamefont {Bian}},
  \bibinfo {author} {\bibfnamefont {M.}~\bibnamefont {Neupane}}, \emph
  {et~al.},\ }\bibfield  {title} {\bibinfo {title} {New type of weyl semimetal
  with quadratic double weyl fermions},\ }\href@noop {} {\bibfield  {journal}
  {\bibinfo  {journal} {Proceedings of the National Academy of Sciences}\
  }\textbf {\bibinfo {volume} {113}},\ \bibinfo {pages} {1180} (\bibinfo {year}
  {2016}{\natexlab{b}})}\BibitemShut {NoStop}%
\bibitem [{\citenamefont {Takane}\ \emph {et~al.}(2019)\citenamefont {Takane},
  \citenamefont {Wang}, \citenamefont {Souma}, \citenamefont {Nakayama},
  \citenamefont {Nakamura}, \citenamefont {Oinuma}, \citenamefont {Nakata},
  \citenamefont {Iwasawa}, \citenamefont {Cacho}, \citenamefont {Kim} \emph
  {et~al.}}]{takane2019observation}%
  \BibitemOpen
  \bibfield  {author} {\bibinfo {author} {\bibfnamefont {D.}~\bibnamefont
  {Takane}}, \bibinfo {author} {\bibfnamefont {Z.}~\bibnamefont {Wang}},
  \bibinfo {author} {\bibfnamefont {S.}~\bibnamefont {Souma}}, \bibinfo
  {author} {\bibfnamefont {K.}~\bibnamefont {Nakayama}}, \bibinfo {author}
  {\bibfnamefont {T.}~\bibnamefont {Nakamura}}, \bibinfo {author}
  {\bibfnamefont {H.}~\bibnamefont {Oinuma}}, \bibinfo {author} {\bibfnamefont
  {Y.}~\bibnamefont {Nakata}}, \bibinfo {author} {\bibfnamefont
  {H.}~\bibnamefont {Iwasawa}}, \bibinfo {author} {\bibfnamefont
  {C.}~\bibnamefont {Cacho}}, \bibinfo {author} {\bibfnamefont
  {T.}~\bibnamefont {Kim}}, \emph {et~al.},\ }\bibfield  {title} {\bibinfo
  {title} {Observation of chiral fermions with a large topological charge and
  associated fermi-arc surface states in cosi},\ }\href@noop {} {\bibfield
  {journal} {\bibinfo  {journal} {Physical review letters}\ }\textbf {\bibinfo
  {volume} {122}},\ \bibinfo {pages} {076402} (\bibinfo {year}
  {2019})}\BibitemShut {NoStop}%
\bibitem [{\citenamefont {Lu}\ \emph {et~al.}(2014)\citenamefont {Lu},
  \citenamefont {Joannopoulos},\ and\ \citenamefont
  {Solja{\v{c}}i{\'c}}}]{lu2014topological}%
  \BibitemOpen
  \bibfield  {author} {\bibinfo {author} {\bibfnamefont {L.}~\bibnamefont
  {Lu}}, \bibinfo {author} {\bibfnamefont {J.~D.}\ \bibnamefont
  {Joannopoulos}},\ and\ \bibinfo {author} {\bibfnamefont {M.}~\bibnamefont
  {Solja{\v{c}}i{\'c}}},\ }\bibfield  {title} {\bibinfo {title} {Topological
  photonics},\ }\href@noop {} {\bibfield  {journal} {\bibinfo  {journal}
  {Nature photonics}\ }\textbf {\bibinfo {volume} {8}},\ \bibinfo {pages} {821}
  (\bibinfo {year} {2014})}\BibitemShut {NoStop}%
\bibitem [{\citenamefont {Shastri}\ \emph {et~al.}(2017)\citenamefont
  {Shastri}, \citenamefont {Yang},\ and\ \citenamefont
  {Zhang}}]{shastri2017realizing}%
  \BibitemOpen
  \bibfield  {author} {\bibinfo {author} {\bibfnamefont {K.}~\bibnamefont
  {Shastri}}, \bibinfo {author} {\bibfnamefont {Z.}~\bibnamefont {Yang}},\ and\
  \bibinfo {author} {\bibfnamefont {B.}~\bibnamefont {Zhang}},\ }\bibfield
  {title} {\bibinfo {title} {Realizing type-ii weyl points in an optical
  lattice},\ }\href@noop {} {\bibfield  {journal} {\bibinfo  {journal}
  {Physical Review B}\ }\textbf {\bibinfo {volume} {95}},\ \bibinfo {pages}
  {014306} (\bibinfo {year} {2017})}\BibitemShut {NoStop}%
\bibitem [{\citenamefont {Yang}\ \emph {et~al.}(2018)\citenamefont {Yang},
  \citenamefont {Guo}, \citenamefont {Tremain}, \citenamefont {Liu},
  \citenamefont {Barr}, \citenamefont {Yan}, \citenamefont {Gao}, \citenamefont
  {Liu}, \citenamefont {Xiang}, \citenamefont {Chen} \emph
  {et~al.}}]{yang2018ideal}%
  \BibitemOpen
  \bibfield  {author} {\bibinfo {author} {\bibfnamefont {B.}~\bibnamefont
  {Yang}}, \bibinfo {author} {\bibfnamefont {Q.}~\bibnamefont {Guo}}, \bibinfo
  {author} {\bibfnamefont {B.}~\bibnamefont {Tremain}}, \bibinfo {author}
  {\bibfnamefont {R.}~\bibnamefont {Liu}}, \bibinfo {author} {\bibfnamefont
  {L.~E.}\ \bibnamefont {Barr}}, \bibinfo {author} {\bibfnamefont
  {Q.}~\bibnamefont {Yan}}, \bibinfo {author} {\bibfnamefont {W.}~\bibnamefont
  {Gao}}, \bibinfo {author} {\bibfnamefont {H.}~\bibnamefont {Liu}}, \bibinfo
  {author} {\bibfnamefont {Y.}~\bibnamefont {Xiang}}, \bibinfo {author}
  {\bibfnamefont {J.}~\bibnamefont {Chen}}, \emph {et~al.},\ }\bibfield
  {title} {\bibinfo {title} {Ideal weyl points and helicoid surface states in
  artificial photonic crystal structures},\ }\href@noop {} {\bibfield
  {journal} {\bibinfo  {journal} {Science}\ }\textbf {\bibinfo {volume}
  {359}},\ \bibinfo {pages} {1013} (\bibinfo {year} {2018})}\BibitemShut
  {NoStop}%
\bibitem [{\citenamefont {Saba}\ \emph {et~al.}(2017)\citenamefont {Saba},
  \citenamefont {Hamm}, \citenamefont {Baumberg},\ and\ \citenamefont
  {Hess}}]{saba2017group}%
  \BibitemOpen
  \bibfield  {author} {\bibinfo {author} {\bibfnamefont {M.}~\bibnamefont
  {Saba}}, \bibinfo {author} {\bibfnamefont {J.~M.}\ \bibnamefont {Hamm}},
  \bibinfo {author} {\bibfnamefont {J.~J.}\ \bibnamefont {Baumberg}},\ and\
  \bibinfo {author} {\bibfnamefont {O.}~\bibnamefont {Hess}},\ }\bibfield
  {title} {\bibinfo {title} {Group theoretical route to deterministic weyl
  points in chiral photonic lattices},\ }\href@noop {} {\bibfield  {journal}
  {\bibinfo  {journal} {Physical Review Letters}\ }\textbf {\bibinfo {volume}
  {119}},\ \bibinfo {pages} {227401} (\bibinfo {year} {2017})}\BibitemShut
  {NoStop}%
\bibitem [{\citenamefont {Yang}\ \emph {et~al.}(2020)\citenamefont {Yang},
  \citenamefont {Gao}, \citenamefont {Feng}, \citenamefont {Huang},
  \citenamefont {Zhou}, \citenamefont {Yang}, \citenamefont {Chong},\ and\
  \citenamefont {Zhang}}]{yang2020ideal}%
  \BibitemOpen
  \bibfield  {author} {\bibinfo {author} {\bibfnamefont {Y.}~\bibnamefont
  {Yang}}, \bibinfo {author} {\bibfnamefont {Z.}~\bibnamefont {Gao}}, \bibinfo
  {author} {\bibfnamefont {X.}~\bibnamefont {Feng}}, \bibinfo {author}
  {\bibfnamefont {Y.-X.}\ \bibnamefont {Huang}}, \bibinfo {author}
  {\bibfnamefont {P.}~\bibnamefont {Zhou}}, \bibinfo {author} {\bibfnamefont
  {S.~A.}\ \bibnamefont {Yang}}, \bibinfo {author} {\bibfnamefont
  {Y.}~\bibnamefont {Chong}},\ and\ \bibinfo {author} {\bibfnamefont
  {B.}~\bibnamefont {Zhang}},\ }\bibfield  {title} {\bibinfo {title} {Ideal
  unconventional weyl point in a chiral photonic metamaterial},\ }\href@noop {}
  {\bibfield  {journal} {\bibinfo  {journal} {Physical Review Letters}\
  }\textbf {\bibinfo {volume} {125}},\ \bibinfo {pages} {143001} (\bibinfo
  {year} {2020})}\BibitemShut {NoStop}%
\bibitem [{\citenamefont {Li}\ \emph {et~al.}(2018)\citenamefont {Li},
  \citenamefont {Huang}, \citenamefont {Lu}, \citenamefont {Ma},\ and\
  \citenamefont {Liu}}]{li2018weyl}%
  \BibitemOpen
  \bibfield  {author} {\bibinfo {author} {\bibfnamefont {F.}~\bibnamefont
  {Li}}, \bibinfo {author} {\bibfnamefont {X.}~\bibnamefont {Huang}}, \bibinfo
  {author} {\bibfnamefont {J.}~\bibnamefont {Lu}}, \bibinfo {author}
  {\bibfnamefont {J.}~\bibnamefont {Ma}},\ and\ \bibinfo {author}
  {\bibfnamefont {Z.}~\bibnamefont {Liu}},\ }\bibfield  {title} {\bibinfo
  {title} {Weyl points and fermi arcs in a chiral phononic crystal},\
  }\href@noop {} {\bibfield  {journal} {\bibinfo  {journal} {Nature Physics}\
  }\textbf {\bibinfo {volume} {14}},\ \bibinfo {pages} {30} (\bibinfo {year}
  {2018})}\BibitemShut {NoStop}%
\bibitem [{\citenamefont {Ge}\ \emph {et~al.}(2018)\citenamefont {Ge},
  \citenamefont {Ni}, \citenamefont {Tian}, \citenamefont {Gupta},
  \citenamefont {Lu}, \citenamefont {Lin}, \citenamefont {Huang}, \citenamefont
  {Chan},\ and\ \citenamefont {Chen}}]{ge2018experimental}%
  \BibitemOpen
  \bibfield  {author} {\bibinfo {author} {\bibfnamefont {H.}~\bibnamefont
  {Ge}}, \bibinfo {author} {\bibfnamefont {X.}~\bibnamefont {Ni}}, \bibinfo
  {author} {\bibfnamefont {Y.}~\bibnamefont {Tian}}, \bibinfo {author}
  {\bibfnamefont {S.~K.}\ \bibnamefont {Gupta}}, \bibinfo {author}
  {\bibfnamefont {M.-H.}\ \bibnamefont {Lu}}, \bibinfo {author} {\bibfnamefont
  {X.}~\bibnamefont {Lin}}, \bibinfo {author} {\bibfnamefont {W.-D.}\
  \bibnamefont {Huang}}, \bibinfo {author} {\bibfnamefont {C.~T.}\ \bibnamefont
  {Chan}},\ and\ \bibinfo {author} {\bibfnamefont {Y.-F.}\ \bibnamefont
  {Chen}},\ }\bibfield  {title} {\bibinfo {title} {Experimental observation of
  acoustic weyl points and topological surface states},\ }\href@noop {}
  {\bibfield  {journal} {\bibinfo  {journal} {Physical review applied}\
  }\textbf {\bibinfo {volume} {10}},\ \bibinfo {pages} {014017} (\bibinfo
  {year} {2018})}\BibitemShut {NoStop}%
\bibitem [{\citenamefont {Xie}\ \emph {et~al.}(2019)\citenamefont {Xie},
  \citenamefont {Liu}, \citenamefont {Cheng}, \citenamefont {Liu},
  \citenamefont {Chen},\ and\ \citenamefont {Tian}}]{xie2019experimental}%
  \BibitemOpen
  \bibfield  {author} {\bibinfo {author} {\bibfnamefont {B.}~\bibnamefont
  {Xie}}, \bibinfo {author} {\bibfnamefont {H.}~\bibnamefont {Liu}}, \bibinfo
  {author} {\bibfnamefont {H.}~\bibnamefont {Cheng}}, \bibinfo {author}
  {\bibfnamefont {Z.}~\bibnamefont {Liu}}, \bibinfo {author} {\bibfnamefont
  {S.}~\bibnamefont {Chen}},\ and\ \bibinfo {author} {\bibfnamefont
  {J.}~\bibnamefont {Tian}},\ }\bibfield  {title} {\bibinfo {title}
  {Experimental realization of type-ii weyl points and fermi arcs in phononic
  crystal},\ }\href@noop {} {\bibfield  {journal} {\bibinfo  {journal}
  {Physical review letters}\ }\textbf {\bibinfo {volume} {122}},\ \bibinfo
  {pages} {104302} (\bibinfo {year} {2019})}\BibitemShut {NoStop}%
\bibitem [{\citenamefont {Huang}\ \emph {et~al.}(2020)\citenamefont {Huang},
  \citenamefont {Deng}, \citenamefont {Li}, \citenamefont {Lu},\ and\
  \citenamefont {Liu}}]{huang2020ideal}%
  \BibitemOpen
  \bibfield  {author} {\bibinfo {author} {\bibfnamefont {X.}~\bibnamefont
  {Huang}}, \bibinfo {author} {\bibfnamefont {W.}~\bibnamefont {Deng}},
  \bibinfo {author} {\bibfnamefont {F.}~\bibnamefont {Li}}, \bibinfo {author}
  {\bibfnamefont {J.}~\bibnamefont {Lu}},\ and\ \bibinfo {author}
  {\bibfnamefont {Z.}~\bibnamefont {Liu}},\ }\bibfield  {title} {\bibinfo
  {title} {Ideal type-ii weyl phase and topological transition in phononic
  crystals},\ }\href@noop {} {\bibfield  {journal} {\bibinfo  {journal}
  {Physical Review Letters}\ }\textbf {\bibinfo {volume} {124}},\ \bibinfo
  {pages} {206802} (\bibinfo {year} {2020})}\BibitemShut {NoStop}%
\bibitem [{\citenamefont {He}\ \emph {et~al.}(2020)\citenamefont {He},
  \citenamefont {Qiu}, \citenamefont {Cai}, \citenamefont {Xiao}, \citenamefont
  {Ke}, \citenamefont {Zhang},\ and\ \citenamefont {Liu}}]{he2020observation}%
  \BibitemOpen
  \bibfield  {author} {\bibinfo {author} {\bibfnamefont {H.}~\bibnamefont
  {He}}, \bibinfo {author} {\bibfnamefont {C.}~\bibnamefont {Qiu}}, \bibinfo
  {author} {\bibfnamefont {X.}~\bibnamefont {Cai}}, \bibinfo {author}
  {\bibfnamefont {M.}~\bibnamefont {Xiao}}, \bibinfo {author} {\bibfnamefont
  {M.}~\bibnamefont {Ke}}, \bibinfo {author} {\bibfnamefont {F.}~\bibnamefont
  {Zhang}},\ and\ \bibinfo {author} {\bibfnamefont {Z.}~\bibnamefont {Liu}},\
  }\bibfield  {title} {\bibinfo {title} {Observation of quadratic weyl points
  and double-helicoid arcs},\ }\href@noop {} {\bibfield  {journal} {\bibinfo
  {journal} {Nature communications}\ }\textbf {\bibinfo {volume} {11}},\
  \bibinfo {pages} {1} (\bibinfo {year} {2020})}\BibitemShut {NoStop}%
\bibitem [{\citenamefont {Yang}\ \emph {et~al.}(2019)\citenamefont {Yang},
  \citenamefont {Sun}, \citenamefont {Xia}, \citenamefont {Xue}, \citenamefont
  {Gao}, \citenamefont {Ge}, \citenamefont {Jia}, \citenamefont {Yuan},
  \citenamefont {Chong},\ and\ \citenamefont {Zhang}}]{yang2019topological}%
  \BibitemOpen
  \bibfield  {author} {\bibinfo {author} {\bibfnamefont {Y.}~\bibnamefont
  {Yang}}, \bibinfo {author} {\bibfnamefont {H.-x.}\ \bibnamefont {Sun}},
  \bibinfo {author} {\bibfnamefont {J.-p.}\ \bibnamefont {Xia}}, \bibinfo
  {author} {\bibfnamefont {H.}~\bibnamefont {Xue}}, \bibinfo {author}
  {\bibfnamefont {Z.}~\bibnamefont {Gao}}, \bibinfo {author} {\bibfnamefont
  {Y.}~\bibnamefont {Ge}}, \bibinfo {author} {\bibfnamefont {D.}~\bibnamefont
  {Jia}}, \bibinfo {author} {\bibfnamefont {S.-q.}\ \bibnamefont {Yuan}},
  \bibinfo {author} {\bibfnamefont {Y.}~\bibnamefont {Chong}},\ and\ \bibinfo
  {author} {\bibfnamefont {B.}~\bibnamefont {Zhang}},\ }\bibfield  {title}
  {\bibinfo {title} {Topological triply degenerate point with double fermi
  arcs},\ }\href@noop {} {\bibfield  {journal} {\bibinfo  {journal} {Nature
  Physics}\ }\textbf {\bibinfo {volume} {15}},\ \bibinfo {pages} {645}
  (\bibinfo {year} {2019})}\BibitemShut {NoStop}%
\bibitem [{\citenamefont {He}\ \emph {et~al.}(2018)\citenamefont {He},
  \citenamefont {Qiu}, \citenamefont {Ye}, \citenamefont {Cai}, \citenamefont
  {Fan}, \citenamefont {Ke}, \citenamefont {Zhang},\ and\ \citenamefont
  {Liu}}]{he2018topological}%
  \BibitemOpen
  \bibfield  {author} {\bibinfo {author} {\bibfnamefont {H.}~\bibnamefont
  {He}}, \bibinfo {author} {\bibfnamefont {C.}~\bibnamefont {Qiu}}, \bibinfo
  {author} {\bibfnamefont {L.}~\bibnamefont {Ye}}, \bibinfo {author}
  {\bibfnamefont {X.}~\bibnamefont {Cai}}, \bibinfo {author} {\bibfnamefont
  {X.}~\bibnamefont {Fan}}, \bibinfo {author} {\bibfnamefont {M.}~\bibnamefont
  {Ke}}, \bibinfo {author} {\bibfnamefont {F.}~\bibnamefont {Zhang}},\ and\
  \bibinfo {author} {\bibfnamefont {Z.}~\bibnamefont {Liu}},\ }\bibfield
  {title} {\bibinfo {title} {Topological negative refraction of surface
  acoustic waves in a weyl phononic crystal},\ }\href@noop {} {\bibfield
  {journal} {\bibinfo  {journal} {Nature}\ }\textbf {\bibinfo {volume} {560}},\
  \bibinfo {pages} {61} (\bibinfo {year} {2018})}\BibitemShut {NoStop}%
\bibitem [{\citenamefont {Peri}\ \emph {et~al.}(2019)\citenamefont {Peri},
  \citenamefont {Serra-Garcia}, \citenamefont {Ilan},\ and\ \citenamefont
  {Huber}}]{peri2019axial}%
  \BibitemOpen
  \bibfield  {author} {\bibinfo {author} {\bibfnamefont {V.}~\bibnamefont
  {Peri}}, \bibinfo {author} {\bibfnamefont {M.}~\bibnamefont {Serra-Garcia}},
  \bibinfo {author} {\bibfnamefont {R.}~\bibnamefont {Ilan}},\ and\ \bibinfo
  {author} {\bibfnamefont {S.~D.}\ \bibnamefont {Huber}},\ }\bibfield  {title}
  {\bibinfo {title} {Axial-field-induced chiral channels in an acoustic weyl
  system},\ }\href@noop {} {\bibfield  {journal} {\bibinfo  {journal} {Nature
  Physics}\ }\textbf {\bibinfo {volume} {15}},\ \bibinfo {pages} {357}
  (\bibinfo {year} {2019})}\BibitemShut {NoStop}%
\bibitem [{\citenamefont {Lee}\ \emph {et~al.}(2018)\citenamefont {Lee},
  \citenamefont {Imhof}, \citenamefont {Berger}, \citenamefont {Bayer},
  \citenamefont {Brehm}, \citenamefont {Molenkamp}, \citenamefont {Kiessling},\
  and\ \citenamefont {Thomale}}]{lee2018topolectrical}%
  \BibitemOpen
  \bibfield  {author} {\bibinfo {author} {\bibfnamefont {C.~H.}\ \bibnamefont
  {Lee}}, \bibinfo {author} {\bibfnamefont {S.}~\bibnamefont {Imhof}}, \bibinfo
  {author} {\bibfnamefont {C.}~\bibnamefont {Berger}}, \bibinfo {author}
  {\bibfnamefont {F.}~\bibnamefont {Bayer}}, \bibinfo {author} {\bibfnamefont
  {J.}~\bibnamefont {Brehm}}, \bibinfo {author} {\bibfnamefont {L.~W.}\
  \bibnamefont {Molenkamp}}, \bibinfo {author} {\bibfnamefont {T.}~\bibnamefont
  {Kiessling}},\ and\ \bibinfo {author} {\bibfnamefont {R.}~\bibnamefont
  {Thomale}},\ }\bibfield  {title} {\bibinfo {title} {Topolectrical circuits},\
  }\href@noop {} {\bibfield  {journal} {\bibinfo  {journal} {Communications
  Physics}\ }\textbf {\bibinfo {volume} {1}},\ \bibinfo {pages} {1} (\bibinfo
  {year} {2018})}\BibitemShut {NoStop}%
\bibitem [{\citenamefont {Li}\ \emph {et~al.}(2021{\natexlab{b}})\citenamefont
  {Li}, \citenamefont {Lv}, \citenamefont {Tao}, \citenamefont {Shi},
  \citenamefont {Chong}, \citenamefont {Zhang},\ and\ \citenamefont
  {Chen}}]{li2021ideal}%
  \BibitemOpen
  \bibfield  {author} {\bibinfo {author} {\bibfnamefont {R.}~\bibnamefont
  {Li}}, \bibinfo {author} {\bibfnamefont {B.}~\bibnamefont {Lv}}, \bibinfo
  {author} {\bibfnamefont {H.}~\bibnamefont {Tao}}, \bibinfo {author}
  {\bibfnamefont {J.}~\bibnamefont {Shi}}, \bibinfo {author} {\bibfnamefont
  {Y.}~\bibnamefont {Chong}}, \bibinfo {author} {\bibfnamefont
  {B.}~\bibnamefont {Zhang}},\ and\ \bibinfo {author} {\bibfnamefont
  {H.}~\bibnamefont {Chen}},\ }\bibfield  {title} {\bibinfo {title} {Ideal
  type-ii weyl points in topological circuits},\ }\href@noop {} {\bibfield
  {journal} {\bibinfo  {journal} {National science review}\ }\textbf {\bibinfo
  {volume} {8}},\ \bibinfo {pages} {nwaa192} (\bibinfo {year}
  {2021}{\natexlab{b}})}\BibitemShut {NoStop}%
\bibitem [{\citenamefont {Wang}\ \emph {et~al.}(2022)\citenamefont {Wang},
  \citenamefont {Zhou}, \citenamefont {Zhang}, \citenamefont {Wu},
  \citenamefont {Yu},\ and\ \citenamefont {Yang}}]{wang2022single}%
  \BibitemOpen
  \bibfield  {author} {\bibinfo {author} {\bibfnamefont {X.}~\bibnamefont
  {Wang}}, \bibinfo {author} {\bibfnamefont {F.}~\bibnamefont {Zhou}}, \bibinfo
  {author} {\bibfnamefont {Z.}~\bibnamefont {Zhang}}, \bibinfo {author}
  {\bibfnamefont {W.}~\bibnamefont {Wu}}, \bibinfo {author} {\bibfnamefont
  {Z.-M.}\ \bibnamefont {Yu}},\ and\ \bibinfo {author} {\bibfnamefont {S.~A.}\
  \bibnamefont {Yang}},\ }\bibfield  {title} {\bibinfo {title} {Single pair of
  weyl points in nonmagnetic crystals},\ }\href@noop {} {\bibfield  {journal}
  {\bibinfo  {journal} {arXiv:2203.13974}\ } (\bibinfo {year}
  {2022})}\BibitemShut {NoStop}%
\bibitem [{sm()}]{sm}%
  \BibitemOpen
  \href@noop {} {\bibinfo  {journal} {See Supplemental Material for the
  tight-binding model, computational and experimental methods}\ }\BibitemShut
  {NoStop}%
\bibitem [{\citenamefont {Hatsugai}(1993)}]{hatsugai1993chern}%
  \BibitemOpen
\bibfield  {journal} {  }\bibfield  {author} {\bibinfo {author} {\bibfnamefont
  {Y.}~\bibnamefont {Hatsugai}},\ }\bibfield  {title} {\bibinfo {title} {Chern
  number and edge states in the integer quantum hall effect},\ }\href@noop {}
  {\bibfield  {journal} {\bibinfo  {journal} {Physical review letters}\
  }\textbf {\bibinfo {volume} {71}},\ \bibinfo {pages} {3697} (\bibinfo {year}
  {1993})}\BibitemShut {NoStop}%
\bibitem [{\citenamefont {Ryu}\ and\ \citenamefont
  {Hatsugai}(2002)}]{ryu2002topological}%
  \BibitemOpen
  \bibfield  {author} {\bibinfo {author} {\bibfnamefont {S.}~\bibnamefont
  {Ryu}}\ and\ \bibinfo {author} {\bibfnamefont {Y.}~\bibnamefont {Hatsugai}},\
  }\bibfield  {title} {\bibinfo {title} {Topological origin of zero-energy edge
  states in particle-hole symmetric systems},\ }\href@noop {} {\bibfield
  {journal} {\bibinfo  {journal} {Physical review letters}\ }\textbf {\bibinfo
  {volume} {89}},\ \bibinfo {pages} {077002} (\bibinfo {year}
  {2002})}\BibitemShut {NoStop}%
\end{thebibliography}%

\end{document}